\documentclass[longauth]{aa}

\newcommand{\hb}{H$\beta$}
\newcommand{\ha}{H$\alpha$}

\newcommand{\oiii}{[O\,{\footnotesize III}]}
\newcommand{\oii}{[O\,{\footnotesize II}]}
\newcommand{\nii}{[N\,{\footnotesize II}]}

\newcommand{\kms}{km s$^{-1}$}
\newcommand{\myr}{$M_{\odot}$ yr$^{-1}$}

\newcommand{\dn}{$D_n 4000$}

\usepackage{graphicx}
\usepackage{txfonts}
\usepackage{lipsum}
\usepackage{subcaption}         % necessary for continued figures, example in section 3
\usepackage{lscape}             % to rotate a single page table, example in appendix.
\usepackage{placeins}           % useful with \FloatBarrier, to keep 
\usepackage{booktabs}
\usepackage[dvipsnames]{xcolor}
\usepackage[colorlinks=true, citecolor=blue]{hyperref}

\begin{document}

\title{DeepDive: A deep dive into the physics of the first massive quiescent galaxies in the Universe}

\author{K.~Ito\inst{1,2}
\and F.~Valentino\inst{1,2}
\and G.~Brammer\inst{1,3}
\and M.~L.~Hamadouche\inst{4}
\and K.~E.~Whitaker\inst{4,1}
\and V.~Kokorev\inst{5}
\and P.~Zhu\inst{1,2}
\and T.~Kakimoto\inst{6,7}
\and P.-F.~Wu\inst{8}
\and J.~Antwi-Danso\inst{9}
\and W.~M.~Baker\inst{10}
\and D.~Ceverino\inst{11,12}
\and A.~L.~Faisst\inst{13}
\and M.~Farcy\inst{14}
\and S.~Fujimoto\inst{9,15}
\and A.~Gallazzi\inst{16}
\and S.~Gillman\inst{1,2}
\and R.~Gottumukkala\inst{1,3}
\and K.~E.~Heintz\inst{1,3,17}
\and M.~Hirschmann\inst{14}
\and C.~K.~Jespersen\inst{18}
\and M.~Kubo\inst{19}
\and M.~Lee\inst{1,2}
\and G.~Magdis\inst{1,2,3}
\and M.~Onodera\inst{6,20}
\and R.~Shimakawa\inst{21}
\and M.~Tanaka\inst{6,7}
\and S.~Toft\inst{1,3}
\and J.~R.~Weaver\inst{4}}

\institute{Cosmic Dawn Center (DAWN), Copenhagen, Denmark
\and 
DTU Space, Technical University of Denmark, Elektrovej 327, DK2800 Kgs. Lyngby, Denmark
\and
Niels Bohr Institute, University of Copenhagen, Jagtvej 128, 2200 Copenhagen N, Denmark
\and
Department of Astronomy, University of Massachusetts, Amherst, MA 01003, USA
\and
Department of Astronomy, The University of Texas at Austin, Austin, TX 78712, USA
\and
Department of Astronomical Science, The Graduate University for Advanced Studies, SOKENDAI, 2-21-1 Osawa, Mitaka, Tokyo 181-8588, Japan
\and 
National Astronomical Observatory of Japan, 2-21-1 Osawa, Mitaka, Tokyo 181-8588, Japan
\and 
Institute of Astrophysics, National Taiwan University, Taipei 10617, Taiwan
\and
David A. Dunlap Department of Astronomy and Astrophysics, University of Toronto, 50 St. George Street, Toronto, Ontario, M5S 3H4, Canada
\and
DARK, Niels Bohr Institute, University of Copenhagen, Jagtvej 155A, DK-2200 Copenhagen, Denmark 
\and 
Departamento de Fisica Teorica, Modulo 8, Facultad de Ciencias, Universidad Autonoma de Madrid, 28049 Madrid, Spain
\and 
CIAFF, Facultad de Ciencias, Universidad Autonoma de Madrid, 28049 Madrid, Spain
\and
Caltech/IPAC, MS 314-6, 1200 E. California Blvd. Pasadena, CA 91125, USA
\and    
\'Ecole Polytechnique F\'ed\'erale de Lausanne (EPFL), Observatoire de Sauverny, Chemin Pegasi 51, CH-1290 Versoix, Switzerland 
\and 
Dunlap Institute for Astronomy and Astrophysics, 50 St. George Street, Toronto, Ontario, M5S 3H4, Canada
\and
INAF - Osservatorio Astrofisico di Arcetri, Largo Enrico Fermi 5, 50125 Firenze, Italy
\and
Department of Astronomy, University of Geneva, Chemin Pegasi 51, 1290 Versoix, Switzerland
\and
Department of Astrophysical Sciences, Princeton University, Princeton, NJ 08544, USA
\and
Department of Physics and Astronomy, School of Science, Kwansei Gakuin University, 2-1 Gakuen, Sanda, Hyogo 669-1337, Japan
\and
Subaru Telescope, National Astronomical Observatory of Japan, National Institutes of Natural Sciences (NINS), 650 North A'ohoku Place, Hilo, HI 96720, USA
\and
Waseda Institute for Advanced Study (WIAS), Waseda University, 1-21-1 Nishi-Waseda, Shinjuku, Tokyo 169-0051, Japan}
   
 \date{Received -, Accepted -} 
  \abstract{We present the DeepDive program, in which we obtained deep ($1-3$ hours) JWST/NIRSpec G235M/F170LP spectra for ten primary massive ($\log{(M_\star/M_\odot)}=10.8-11.5$) quiescent galaxies at $z\sim3-4$. A novel reduction procedure was used to extend the nominal wavelength coverage of G235M beyond \ha\ and \nii\ at $z\sim 4$, revealing weak, narrow \ha\ lines indicative of low star formation rates (${\rm SFR}\sim0-5\, M_\odot\, {\rm yr^{-1}}$). Two out of ten primary targets have broad \ha\ lines, indicating the presence of active galactic nuclei. We also conducted an archival search of quiescent galaxies observed with NIRSpec gratings in the DAWN JWST Archive, which provides a statistical context for interpreting the DeepDive targets. This archival search provided a spectroscopic sample of 126 quiescent galaxies spanning $1<z<5$, selected by high \dn, UVJ color, or low specific star formation rate (more than one dex lower than that of the star formation main sequence), and covering more than an order of magnitude in stellar mass. This sample allowed us to revisit the sample from the different selections, finding $\sim90\%$ overlap between these criteria. The total sample of 136 quiescent galaxies constructed in this study shows that those at $z\sim3-5$, including the DeepDive targets, typically exhibit weaker 4000 \AA\ breaks and bluer colors than their lower-redshift counterparts, indicating generally younger stellar populations. Stacked spectra of sources grouped by the \dn\ index reveal faint iron and magnesium absorption line features in the stellar continuum even for the low \dn\ (\dn $<1.35$) subsample at high redshift ($z\sim3$). In addition, higher \dn\ subsamples show fainter nebular emission lines. These results demonstrate that medium-resolution NIRSpec spectroscopy is essential for robustly characterizing the diversity and evolution of early quiescent galaxies. The large sample constructed in this paper will allow a statistical census of the properties of quiescent galaxies at high redshift to be obtained. All photometric and spectroscopic data from this study have been made publicly available.}

   \maketitle
   \nolinenumbers

\section{Introduction}
\noindent
Over the past two decades, a surge of studies has shed light on several aspects of the formation and evolution of the first massive quiescent galaxies, characterized by early and intense star formation activity followed by sudden ``quenching'' and then a long passive evolution of their stellar populations. The concept of ``first'' quiescent systems has been revisited each time new and more powerful instruments have become available. Only 20 years ago, the numerous population of red and morphologically compact galaxies at $z\sim1.5-2$ was indicated as the prototype of early massive quiescent galaxies \citep[e.g.,][]{daddi_2005, trujillo_2006, toft_2007}, later confirmed when the first near-infrared spectrographs installed on 8-10m ground-based telescopes were built \citep{kriek_2006,vandokkum_2009, kriek_2009}. Spectroscopy has been instrumental in confirming the redshifts of massive quiescent galaxies, establishing the existence of old stellar populations, constraining their star formation histories, and estimating their dynamical masses \citep{belli_2014, belli_2017a}. Since then, the quiescent frontier has been pushed to $z\sim3-4$, the limit at which the brightest ($K\lesssim22-23\, {\rm mag}$) and most massive systems could be spectroscopically confirmed in the last transparent window in the near-infrared offered by the atmosphere \citep[e.g.,][]{glazebrook_2017, schreiber_2018c, tanaka_2019, valentino_2020a, forrest_2020b, forrest_2022, esdaile_2021, kakimoto_2024, antwi-danso_2025}. Recently, the advent of the James Webb Space Telescope (JWST) has allowed for a spectacular jump forward in the exploration of this topic. Massive systems with suppressed star formation rates (SFRs) have been spectroscopically confirmed up to $z\sim5-7$, posing a serious challenge to our models and simulations \citep{carnall_2024,deGraaff_2024,onoue_2024,lagos_2024,delucia_2024,weibel_2025}. The unprecedented sensitivity of JWST has revealed the presence of pockets of outflowing ionized and neutral gas, possibly powered by residual star formation or active galactic nuclei (AGNs) activity \citep{belli_2024,davies_2024,d'eugenio_2023,wu_2024,valentino_2025}, the standard agents invoked to explain the abrupt interruption of intense star formation \citep[][for recent overviews]{lagos_2024,delucia_2025}. Tighter constraints on the star formation history (SFHs), showing the existence of short star formation periods \citep[e.g.,][]{carnall_2024,park_2024,baker_2024,nanayakkara_2025}, and, at lower redshifts, on the history of metal enrichment \citep[e.g.,][]{gallazzi_2014,beverage_2021,beverage_2024,beverage_2025} can now constitute excellent tests for the next-generation hydrodynamical simulations and have prompted the creation of ever novel spectral models to reproduce and explain the observed spectra \citep[e.g.,][]{Park_2024_alphamc}.

Throughout this exploration of quiescent galaxies, research has typically proceeded in two phases: an initial ``discovery'' phase, followed by ``population studies.'' The early phase focuses on identifying remarkable systems, while the later stage aims to characterize galaxies on a statistical basis. Since the launch of JWST, the emphasis has been on finding exceptional systems -- the most distant, most massive, or oldest galaxies -- to demonstrate the capabilities of the instruments and push the limits of prior knowledge. However, only statistical studies can ultimately reveal what is typical versus exceptional and establish representative samples for meaningful comparisons with theoretical models.

In this work, we introduce the ``A deep dive into the physics of the first massive quiescent galaxies in the Universe'' (DeepDive) program, a spectroscopic campaign with JWST/NIRspec aimed at characterizing the properties and underlying physics of massive quiescent galaxies at $z=3-4$. New observations are placed in context by leveraging all equivalent JWST/NIRSpec Micro-Shutter Assembly (MSA) observations taken within the first three years of telescope operations, retrieved from the DAWN JWST Archive (DJA). Our ultimate goal is to provide the basis for the first robust statistical census of this population at $1<z<5$ and to offer a complementary spectroscopic analysis to the photometric archival approach presented in \cite{valentino_2023} and \citet{baker_2025b}. Here we introduce the selection, observations, and data reduction of the DeepDive program (Section~\ref{sec:deepdive}). We then describe our archival search for grating spectra of quiescent galaxies (Section~\ref{sec:archive}) and present their basic properties, such as stellar masses, SFR, X-ray detections, colors, and indicators of the strength of the 4000\AA\ stellar break (Section~\ref{sec:properties}). In the same section, we present average spectra obtained by stacking subsamples grouped by the \dn\ index and explore their properties. The analysis of specific scientific topics, including the histories of star formation and metal enrichment, stellar dynamics and structures, gas outflows, and Interstellar Medium (ISM) properties, is deferred to dedicated forthcoming works (Section \ref{sec:sciencegoal}). All spectroscopic and photometric data will be released in conjunction with this paper, with the aim of accelerating progress in our understanding of how massive quiescent galaxies form and evolve and better informing future time allocation requests. Throughout this work, we make use of the AB system \citep{oke_1983} to report magnitudes. We adopted a $\Lambda$ cold dark matter cosmology with $\Omega_{\rm m} = 0.3$, $\Omega_{\Lambda} = 0.7$, and $H_0 = 70\,\mathrm{km\,s^{-1}\,Mpc^{-1}}$.

\section{The DeepDive program}
\label{sec:deepdive}
The DeepDive program (PID \#3567, PI: Valentino) is a JWST General Observer (GO) program targeting quiescent galaxies at $z\sim3-4$ with the NIRSpec and NIRCam instruments. Its primary goals are to derive fundamental physical properties of these galaxies, such as star formation histories, metallicities, and stellar velocity dispersions, from deep grating spectra and photometry (see Section~\ref{sec:sciencegoal} for more details). In this section, we describe the target selection, observational data, and the procedures for data reduction.

\subsection{Primary target selection}
The primary targets of DeepDive are 11 bright ($K<23$ mag), and therefore massive ($M_\star\sim10^{11}\,M_\odot$), quenched galaxies with known spectroscopic redshifts from ground-based observations at $z=3.4-4$ \citep{glazebrook_2017,schreiber_2018c,schreiber_2018b,tanaka_2019,esdaile_2021,forrest_2020b,valentino_2020a}. These were among the most distant quiescent galaxies known prior to the launch of JWST. One target was not observed due to guiding failure in the NIRCam imaging and was thus excluded from the analysis. The remaining ten objects and their properties are listed in Table \ref{tab:maintargets}. The primary target selection is based on an extension of the rest-frame $UVJ$ colors that includes more recently quenched post-starburst-like objects \citep{williams_2009, schreiber_2018b, belli_2019}. The targets are scattered among different cosmological fields, some of which are only partially covered by the existing JWST and HST imaging previously, as detailed in Section~\ref{sec:imaging}: PRIMER-UDS and COSMOS \citep{donnan_2024}, COSMOS-Web \citep{casey_2023}, CEERS-EGS \citep{bagley_2023,finkelstein_2023}, and XMM-VIDEO \citep{jarvis_2013}. 

\begin{figure*}
    \centering
    \includegraphics[width=0.92\linewidth]{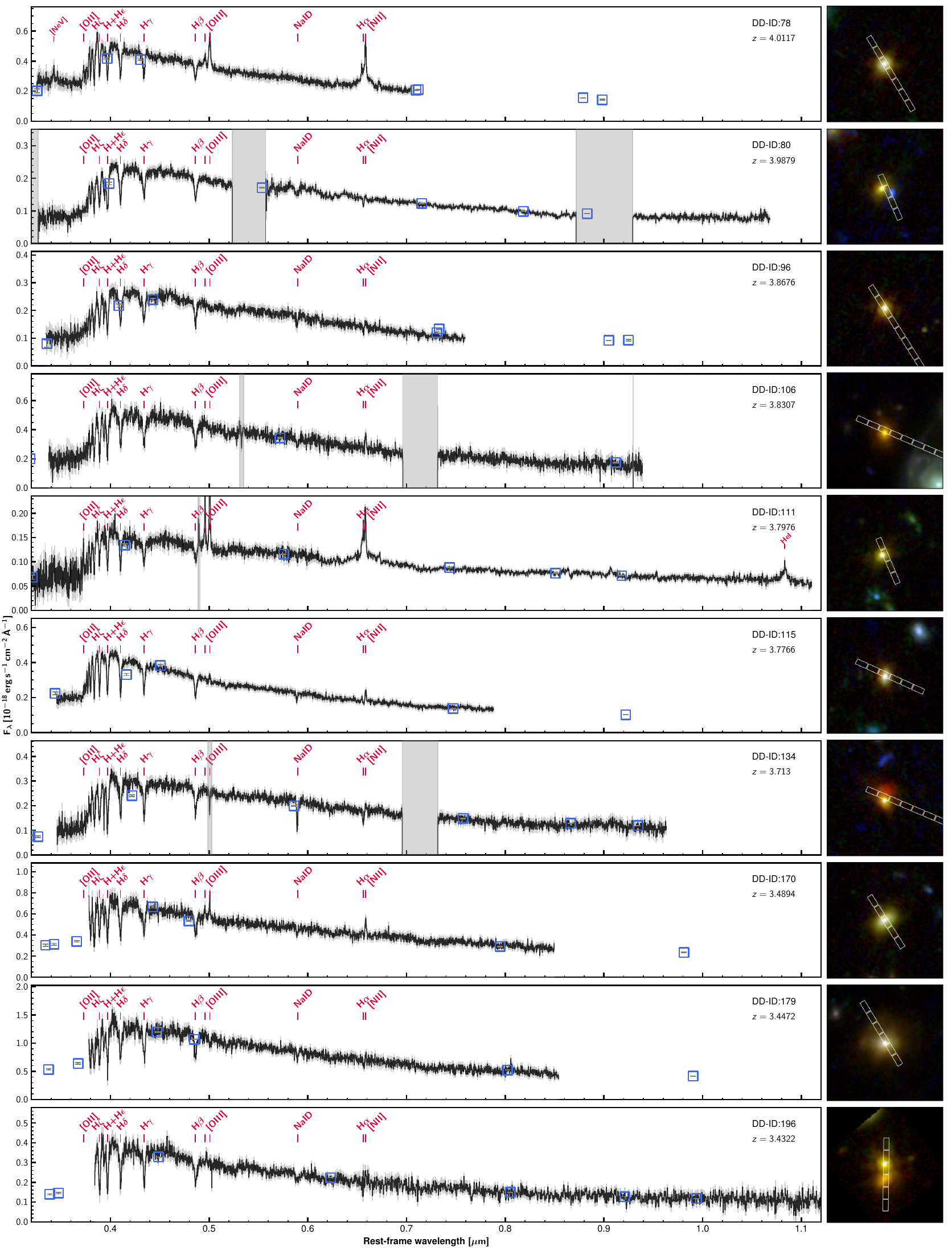}
    \caption{Left: Spectra of the main DeepDive targets. Spectra are shown in solid black lines, and gray hatched regions represent their $1\sigma$ uncertainties. The gray rectangles correspond to the wavelength lost due to the gaps among detectors and masked regions. Blue open squares mark the available photometry. Right: JWST/NIRCam RGB color image of the main targets with the shutter position in one dither pattern. The image size is $5$\arcsec $\times 5$\arcsec. F115W (F150W) images are used for the blue color of IDs 80, 106, 111, 134, and 196 (IDs 78, 96, 115, 170, and 179). F200W (F277W) is used for the green color of ID 78, 80, 96, 111, 115, 134, 170, 179, and 196 (ID 106). F444W images are used for red color.}
    \label{fig:main-spectra}
\end{figure*}

\subsection{Imaging}
\label{sec:imaging}
As part of the DeepDive program, we acquired NIRCam imaging for the five targets previously lacking it (DD-ID\footnote{The ID in the DeepDive program.} 78, 96, 115, 170, 179). We observed each field with four filters (F150W, F200W, F356W, and F444W) for a total integration time of $1,460$s each. We implemented a four-point INTRAMODULE dithering pattern for all targets, except for DD-ID 179, where we used a more compact INTRAMODULEBOX configuration. The observations were split into seven groups, each with one integration, for a total of four integrations per dithering position. The raw data were reduced as described in \cite{valentino_2023}. Briefly, we retrieved level-2 products processed with the JWST pipeline and processed them with \textsc{grizli} \citep{brammer_2023}, where we used the \texttt{jwst\textunderscore0989.pmap} calibration data and the most recent sky flats. All images are aligned to Gaia Data Release 3 \citep{gaia-collaboration_2021}, coadded, and drizzled \citep{fruchter_2002} onto a common pixel scale of $0\farcs04$. This allowed us to reach an average $5\sigma$ depth of 28.3 mag in the F444W image within $0\farcs15$ circular radius apertures (without further aperture corrections). The $1\sigma$ depth was estimated by randomly placing 2000 apertures in source-free regions and computing $1.4826\times\mathrm{MAD}$, where $\mathrm{MAD}$ is the median absolute deviation.

The rest of the targets benefit from existing coverage of JWST/NIRCam and HST imaging from various public programs. The JWST and HST images were retrieved from the DJA\footnote{\url{https://dawn-cph.github.io/dja/index.html}} and processed uniformly according to the steps described above. The Cosmic Evolution Early Release Science Survey \citep[CEERS;][PI: Finkelstein]{bagley_2023,finkelstein_2023} covers with seven NIRCam filters $\sim80$ arcmin$^2$ of the Extended Groth Strip (EGS), a field previously imaged by Hubble as part of the Cosmic Assembly Near-infrared Deep Extragalactic Legacy Survey \citep[CANDELS;][]{grogin_2011,koekemoer_2011} and 3D-HST Survey \citep{brammer_2012}. The NIRCam observations reach a $5\sigma$ sensitivity limit of $28.6$ mag in F444W, extracted within circular apertures of radius $0\farcs1$ and corrected for the aperture losses \citep{bagley_2023}. The Public Release IMaging for Extragalactic Research (PRIMER) Survey (PID \#1837; PI: Dunlop) covers part of two legacy fields, COSMOS ($\sim136$ arcmin$^2$) and UDS ($\sim243$ arcmin$^2$), with eight NIRCam filters. The average depths in the F444W filter reached in the COSMOS and UDS sky patches are of $28.0-28.4$ (wide-deep) and $27.9$ mag for point sources, respectively. These were computed from $0\farcs15$ circular apertures in point spread function (PSF)-matched images, then corrected to total fluxes \citep{donnan_2024}. The COSMOS-Web survey \citep{casey_2023} imaged a larger and contiguous portion of $0.54$ deg$^2$ of the COSMOS field with four NIRCam filters, down to a $5\sigma$ depth of $28.17$ in $0\farcs15$ radius apertures in F444W (without further aperture corrections). Table \ref{tab:photoinfo} summarizes ancillary information on the available JWST and HST photometry for our primary targets.

In addition to the mosaics in their native spatial resolution, we produced a PSF-homogenized version of each image, matched to the PSF of the reddest available NIRCam filter (F444W). The PSF-matched kernels are created using \texttt{Pypher} \citep{boucaud_2016}, where the PSFs are empirical PSFs created from stacking bright, unsaturated stars in each image. For the CEERS and PRIMER fields, we used existing PSF-homogenized images, as documented in the literature \citep[respectively]{wright_2024, cutler_2024}.

Finally, we extracted the photometry in two different ways. Firstly, we followed the standard approach for all photometric catalogs collected on DJA and described in \cite{valentino_2023} and extracted the photometry from the original non-PSF-homogenized images. We ran the pythonic version of the Source Extractor code \citep[\texttt{SEP} v1.2.1,][]{bertin_1996,barbary_2016} to extract sources in a combined image of all available NIRCam LW filters without prior PSF matching. We measured fluxes in circular apertures with a diameter of $0\farcs5$ and corrected them to ``total'' values within Kron apertures \citep{kron_1980}, which were computed on the LW combined detection image. 

Next, we used \textsc{AperPy} to extract robust photometry from the F444W PSF-matched images, following the method described in \cite{weaver_2024_uncover}. \textsc{AperPy} uses \texttt{SEP} to detect sources in a sky-subtracted, noise-equalized (invariance-weighted) detection image based on our deepest long-wavelength \textit{JWST} filters. Photometric measurements are made on the F444W PSF-matched images in 0$^{\prime\prime}$.32, 0$^{\prime\prime}$.48, 0$^{\prime\prime}$.70, 1$^{\prime\prime}$.00 and 1$^{\prime\prime}$.40 diameter circular apertures using \texttt{SEP}. Neighboring sources in the vicinity of the targets were masked out. These measurements were corrected to total fluxes to account for light missed at larger radii by first considering a Kron-like ellipse fit on a PSF-matched version of the detection image. Fluxes were then further corrected to total by adding light in the PSF outside the circularised radius of the Kron-like ellipse (the Kron step is skipped for small, unresolved sources). We refer the reader to \cite{weaver_2024_uncover} for further details on these corrections. Uncertainties on photometric measurements were derived in \textsc{AperPy} by randomly placing circular apertures on empty regions of the PSF-matched detection image, providing an estimate of the background image noise, following the method described in \cite{labbe_2005} and \cite{whitaker_2011}. In summary, total flux errors were calculated by multiplying the noise by the ratio of total to aperture flux for each object. We built a ``super'' catalog which includes the photometry measured in the smallest usable aperture for each filter using isophotal sizes based on the detection image \citep[see][]{labbe_2003,weaver_2024_uncover}. 

The total photometry extracted from the PSF-matched images is systematically brighter than DJA's by $\sim10$\% in every filter without any appreciable color effects (Figure \ref{fig:Mstar-comp}). The systematic offset in brightness is also translated into a similar difference in properties derived from the Spectral Energy Distribution (SED) modeling (Section~\ref{sec:SEDfitting}), such as the stellar mass (Figure \ref{fig:Mstar-comp}). We will use the photometry from PSF-matched images as the fiducial one. However, we release the alternative photometry and scaled spectra (Section~\ref{sec:fluxcorrection}) to allow for a fully homogeneous comparison with the literature sample drawn from DJA (Section~\ref{sec:archive}). 

\begin{figure*}
    \centering
    \includegraphics[width=1\linewidth]{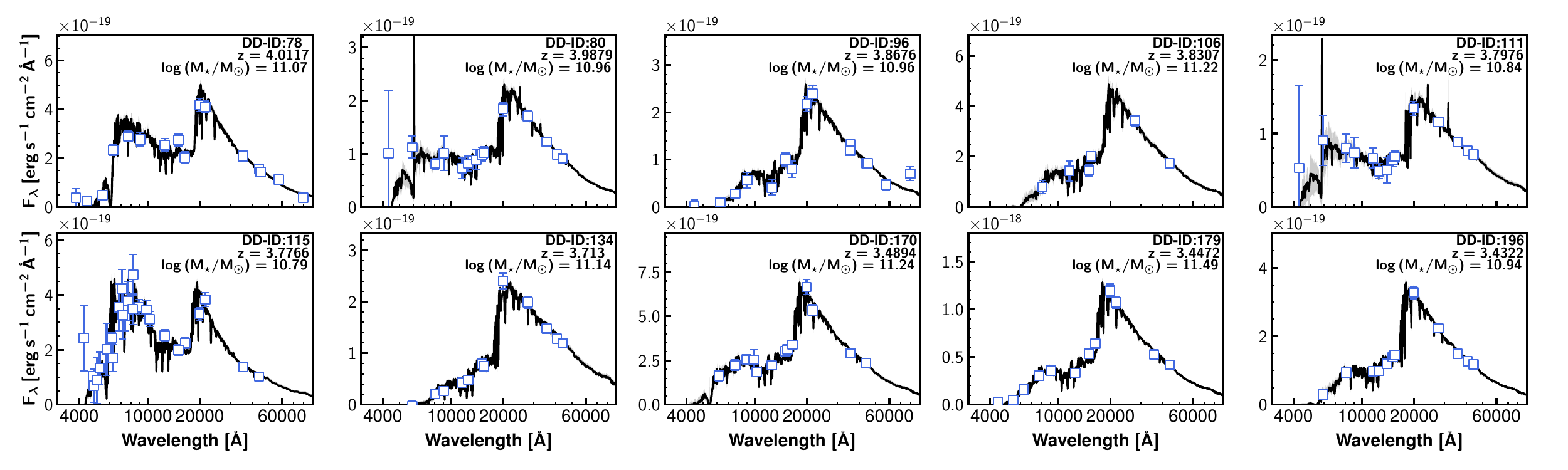}
    \caption{Spectral energy distributions of the main DeepDive targets. Blue squares represent the available photometric data, and black lines show the best-fit SED models obtained using {\sc Bagpipes}. The ID, redshift, and stellar mass of each target are indicated in the corresponding panel.}
    \label{fig:SED-DeepDive}
\end{figure*}

\subsection{Spectroscopy}\label{sec:spectroscopy}
Eight of the ten main targets have been observed in eight JWST/NIRSpec Micro-Shutter Assembly (MSA, \citealt{jakobsen_2022}) pointings as part of the DeepDive program. In each case, we used the G235M/F170LP medium resolution ($R\sim1000$) grating and filter combination to cover wavelengths around the Balmer break for targets at $z>3.4$. Each observation is tailored to the brightness of the main target to ensure a similar S/N ratio on the stellar continuum emission at rest-frame optical wavelengths, resulting in different total integration times. The total integration time ranged from 2,300 s to 10,500 s (Table \ref{tab:photoinfo}). For all masks, we adopted a 3-point nodding pattern. We allocated at least five microshutters to the main targets in order to enclose enough background emission for an optimal subtraction. The rest of the fillers were allocated an average of three microshutters. During the mask preparation phase, we did not allow any spectral overlap between the primary targets and other bright continuum emitters \citep[e.g., dusty star-forming galaxies,][]{Gillman_2026}. However, lower-priority fillers, mostly comprised of line emitters, could have overlapping spectral traces. Here we focus on the main massive, quenched targets. Therefore, the overlap of spectral traces does not constitute an issue. The remaining two main targets (DD-ID 80 and 111) had been previously observed as part of GO PID \#3543 \citep[the Early eXtragalactic Continuum and Emission Line Science, EXCELS, PI: A. Carnall,][]{carnall_2024}. We thus used the raw data from this program and reduced their spectra as described for the rest of the spectra acquired in DeepDive. These targets, common to EXCELS, have been observed for $14,706$s with the G140M and G395M gratings, and for $19,958$s with the G235M grating, respectively. Besides these two targets, we searched for and added coverage of prism and medium resolution gratings from DJA whenever available for our main sample. This resulted in using the PRISM/CLEAR and G395M/F290LP spectra from the GO PID \#4223 \citep[the Red Unknowns: Bright Infrared Extragalactic Survey, RUBIES, PIs: A. de Graaff and G. Brammer,][]{degraaff_2024_rubies} for DD-ID 196 and the PRISM/CLEAR spectrum from GO PID \#2565 (PI: K. Glazebrook) for DD-ID 80 and 134 \citep{nanayakkara_2024, nanayakkara_2025}. All spectra have been homogeneously reduced using {\sc MSAExp} \citep{brammer_2023_msaexp,heintz_2025, degraaff_2024_rubies}. The pipeline also determines the spectroscopic redshifts of the sources. It has been recently updated to recover the full red extension of the spectra allowed by the combination of the long-pass filter and the projection on the detector, as detailed in \cite{valentino_2025} and \citet{pollock_2025} (\textsc{v4})\footnote{\url{https://doi.org/10.5281/zenodo.15472353}}.

\subsection{Flux correction}\label{sec:fluxcorrection} 
We corrected the grating spectra for residual flux losses by anchoring the spectroscopy to the photometry. We applied a polynomial correction function, with an order adapted to the number of available photometric data points in the observed spectral range ($n=0$, $n=1$, and $n=2$ for $1-2$, $3-4$, or $>4$ filters available, respectively). For the correction, we primarily considered the photometry from JWST and HST (Section~\ref{sec:imaging} and Table \ref{tab:photoinfo}). However, due to the limited filter coverage of their fields, we also included the available photometry for DD-IDs 78, 96, 115, 170, and 179 in the literature. For sources in the COSMOS field not covered by the PRIMER survey, the SXDS field, and the XMM-LSS field, we adopted the photometric measurements in the catalogs by \citet{weaver_2022}, \citet{kubo_2018}, and \citet{Desprez_2023}, respectively. We considered the available total photometry in all cases. Sources with DD-IDs 80, 134, and 196 were observed with PRISM/CLEAR as part of other programs \citep[][]{degraaff_2024_rubies,nanayakkara_2024,nanayakkara_2025}. In these cases, we first anchored the prism spectra to the photometry and then the grating spectra to the corrected prism spectra. Possible amplitude differences between prism and grating spectra can arise from different shutter positions. 

Finally, we introduced a second-order wavelength-dependent correction for the portion of the v4 grating spectra extended beyond the nominal cutoff. The correction is necessary to compensate for an imperfect calibration due to the lack of dedicated calibration spectra at certain wavelengths. We derived such a correction by modeling with a spline the ratio between the grating spectra and available prism spectra for a subsample of the DeepDive main targets (DD-ID 80, 134, and 196) and in the archival collection (Section~\ref{sec:archive}), and we applied it to all available medium-resolution spectra. The corrections derived for each grating are shown in Appendix \ref{app:Extendedcorr}. In the case of G235M/F170LP spectra as those collected in DeepDive, the correction amounts to up to 20 percent at $\lambda>3.1\, {\rm \mu m}$. The flux-corrected spectra of all primary targets in our program are shown in Figure \ref{fig:main-spectra}.

\begin{table*}[]
    \small
    \centering
    \caption{Properties of the main targets in DeepDive.}
    \begin{tabular}{lccccccccc}
        \toprule
         DD-ID & (R.A., Decl.) & $z_{\rm spec}$\tablefootmark{a} & $M_\star$\tablefootmark{b} & SFR\tablefootmark{b} &  $A_{V}$\tablefootmark{b} & \dn\tablefootmark{c}\ & ${\rm QG_{D_n 4000}}$\tablefootmark{d} & ${\rm QG_{UVJ}}$\tablefootmark{e} &  ${\rm QG_{sSFR}}$\tablefootmark{f}\\
        & & & [$10^{10}\, M_\odot$] & [$M_\odot\, {\rm yr^{-1}}$] & [mag] & & & &\\
        \midrule
        78 & (34.298697, -4.989901)  & 4.0117 & $11.7 \pm 1.1$ & $0.0 \pm 7.6$ & $0.37 \pm 0.16 $ & $1.102 \pm 0.014$ & 0 & 0 & 1 \\
        80 & (34.340358, -5.241255)  & 3.9879 & $9.1 \pm 0.7$ & $3.9 \pm 2.0$ & $0.11 \pm 0.13 $ & $1.367 \pm 0.017$ & 1 & 1 & 1 \\
        96 & (34.756280, -5.308090)  & 3.8676 & $9.2 \pm 0.6$ & $0.000 \pm 0.087$ & $0.16 \pm 0.11 $ & $1.284 \pm 0.019$ & 1 & 1 & 1 \\
        106 & (149.932856, 2.123422) & 3.8307 & $16.7 \pm 1.4$ & $ 0.00\pm 0.57$ & $0.11 \pm 0.10 $ & $1.324 \pm 0.022$ & 1 & 1 & 1 \\
        111 & (34.289452, -5.269803) & 3.7976 & $7.0 \pm 0.7$ & $8 \pm 21$ & $0.57 \pm 0.27 $ & $1.148 \pm 0.018$ & 0 & 1 & 1 \\
        115 & (149.419594, 2.007526) & 3.7766 & $6.2 \pm 0.3$ & $12 \pm 6$ & $0.04 \pm 0.06 $ & $1.154\pm 0.013$ & 0 & 0 & 1 \\
        134 & (150.061466, 2.378713) & 3.7130 & $13.69 \pm 0.83$ & $0.00 \pm 0.06$ & $0.51 \pm 0.15 $ & $1.385 \pm 0.022$ & 1 & 1 & 1 \\
        170 & (36.733541, -4.536583) & 3.4894 & $17.2 \pm 1.5$ & $0.0 \pm 2.3$ & $0.25 \pm 0.18 $ & $1.20 \pm 0.019$ & 1 & 1 & 1 \\
        179 & (34.386994, -5.482689) & 3.4472 & $31.1 \pm 1.9$ & $0.0 \pm 0.5$ & $0.30 \pm 0.12 $ & $1.346 \pm 0.027$ & 1 & 1 & 1 \\
        196 & (214.866046, 52.884258) & 3.4332 & $8.7 \pm 0.4$ & $0.00 \pm 0.14$ & $0.27 \pm 0.11 $ & $1.262 \pm 0.025$ & 1 & 1 & 1 \\ 
        \bottomrule
    \end{tabular}
    \tablefoot{
        \tablefoottext{a}{The spectroscopic redshift is updated based on the JWST/NIRSpec spectra.}
        \tablefoottext{b}{Quantities obtained by modeling the SED using the photometry extracted from the PSF-matched images.}
    \tablefoottext{c}{\dn\ is measured from the NIRspec spectra.}
    \tablefoottext{d, e, f}{These flags are set to $1$ if a source satisfies the \dn, $UVJ$, and sSFR selection criteria for quiescent galaxies described in Section \ref{sec:archive}, respectively.}
    }
    \label{tab:maintargets}
\end{table*}

\subsection{SED fitting}\label{sec:SEDfitting}
We derived stellar masses, SFRs, dust attenuation, and rest-frame colors of our targets by modeling their SEDs. For the purpose of presenting the general properties of our sample, we opted to model the photometry under common assumptions widely used in the literature for a straightforward comparison. More sophisticated modeling of the combined spectroscopic and photometric data will be reported in a separate work and adapted to the specific science goals \citep{Hamadouche_2026}. Here we used two SED fitting codes, {\sc Eazy-py} \citep{brammer_2008, Brammer_eazy-py_2021} and {\sc Bagpipes} \citep{carnall_2018}. We chose the former to provide estimates of the photometric redshifts and rest-frame $UVJ$ colors that can be immediately compared with photometric samples from the DJA. We used {\sc Bagpipes} to estimate stellar masses and SFRs based on a more standard stellar population modeling approach, which benefits from extensive testing over the past few years \citep[e.g.,][]{carnall_2018,carnall_2019}. We modeled the available JWST and HST photometry extracted from the PSF-matched images, complemented with existing measurements for sources with limited coverage from these telescopes (see Section~\ref{sec:imaging}).

We ran {\sc Eazy-py} in the same configuration as for all photometric catalogs in DJA \citep{valentino_2023}. We used the set of 13 templates from the Flexible Stellar Populations Synthesis code \citep{conroy_2010}, as constructed in \citet{kokorev_2022}\footnote{We used the template set called {\tt agn\_blue\_sfhz\_13.param}.}. We added a template to capture the shape of high-redshift strong line emitters \citep{carnall_2022a}. We adopted a \cite{chabrier_2003} initial mass function (IMF) and the dust attenuation law from \citet{kriek_2013}. In addition, the redshift was fixed to the spectroscopic redshift obtained by running {\sc MSAExp} on the NIRSpec spectra. 

We launched {\sc Bagpipes} with stellar population models from \citet{bruzual_2003} using the MILES stellar spectral library \citep{Sanchez-Blazquez_2006}, stellar evolutionary tracks from \citet{bressan_2012}, and by assuming the IMF from \citet{kroupa_2001}. We did not include any corrections to stellar mass and SFRs induced by the minimal difference with a \citet{chabrier_2003} IMF. The stellar metallicity was set as a free parameter. We assumed a double power-law star formation history model, following the other studies on quiescent galaxies \citep[e.g.,][]{carnall_2024} and applied the dust attenuation law from \citet{calzetti_2000}. Emission line models were included with a fixed ionization parameter of $\log{U}=-3$ \citep{byler_2017}, also following similar studies \citep[e.g.,][]{carnall_2024}. We note that we also fixed the redshift to its spectroscopic estimate in this case. The full list of free parameters, their priors, and limits is summarized in Table \ref{tab:priors}. The best-fit SEDs of the main targets are shown in Figure \ref{fig:SED-DeepDive}. The obtained parameters are summarized in Table \ref{tab:maintargets}.

\subsection{Overview of the spectra and \ha\ SFR measurements} 
Stellar absorption features typical of quenched systems, such as the absorption lines of the Balmer series, are clearly visible in the spectra of all main DeepDive targets (Figure \ref{fig:main-spectra}). All spectra cover wavelengths redder than $\lambda_{\rm rest}>0.38\, {\rm \mu m}$, and thus the $4000$\, \AA\, break and the ionized Calcium CaII H+K lines are detected. Thanks to the extended spectra from the latest v4 reduction, we also recover the wavelength around \ha\ and \nii\ emission lines of objects at $z>3.8$. Interestingly, many sources show several emission lines. We detect the \ha\ emission line with a signal-to-noise ratio $\geq3\ (2)$ in five (seven) out of ten sources. Moreover, such \ha\ emission has a broad component (FWHM$=7000$\, \kms\ and $12000$\, \kms) in 2 out of 10 sources (DD-IDs 78 and 111). This is similar to what has been recently reported in other quenched galaxies at $z=2-5$ \citep{carnall_2023b, ito_2025a, bugiani_2025}. Both DD-ID 78 and 111 are detected in X-ray imaging of the X-UDS \citep{Kocevski_2018}\footnote{The IDs for 78 and 111 in the X-UDS catalog are xuds\_514 and xuds\_039, respectively.}. These suggest that these two galaxies host a type-1 AGN. DD-ID 78 exhibits [NeV]$\lambda3427$, [NeIII]$\lambda3869$, and [OIII]$\lambda4960,\,5008$ emission lines, which further corroborate the presence of an AGN \citep[e.g.,][]{kewley_2019}. 

We then derived an estimate of the instantaneous SFR from the \ha\ emission, assuming that it is solely powered by star formation. We measured the \ha\ flux following the procedure in \citet{ito_2025b}, which conducted the line-fitting of DD-ID 196\footnote{The ID in that paper is 61168.}, the first DeepDive target observed with JWST/NIRSpec. We first modeled the stellar continuum with {\sc ppxf} \citep{cappellari_2017, cappellari_2023}, and then modeled the emission lines as Gaussian functions in the continuum-subtracted spectra around \ha\ and \hb. We simultaneously modeled close emission lines, such as [NII] emission lines, assuming a constant redshift and velocity width. As mentioned, for DD-IDs 78 and 111, the inclusion of a second broad Gaussian component for \ha\ and \hb\ improved the overall fit. 

\begin{figure*}
    \centering
    \includegraphics[width=1\linewidth]{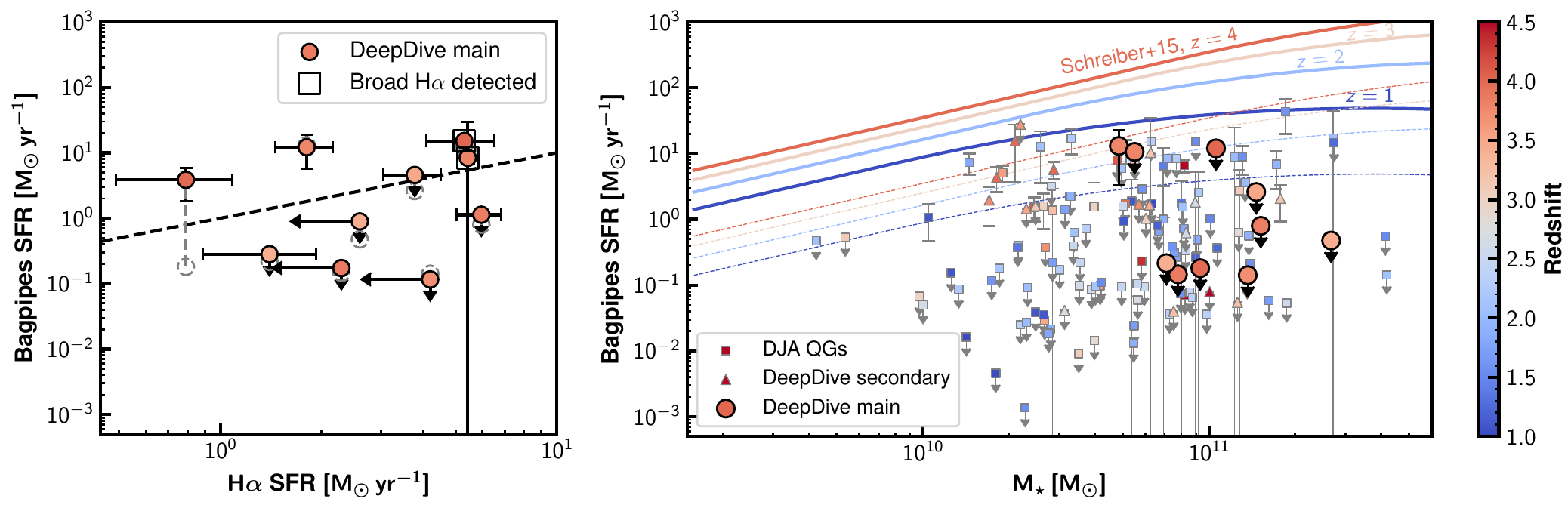}
    \caption{Left panel: Comparison between SFR from the SED fitting and SFR from \ha\ luminosity of the DeepDive main targets. If the SFRs from the SED fitting are below $1\, M_\odot {\rm yr^{-1}}$, they are replaced by a $2\sigma$ upper limit. If the signal-to-noise ratio of \ha\ flux is below two, the SFR from \ha\ luminosity is shown as a $2\sigma$ upper limit. Two sources with broad \ha\ emission are highlighted with squares. The SFRs from the SED fitting using the DJA photometry are shown in dashed circles for easy comparison to the right panel. Right panel: Stellar mass and SFR relation for the ten main targets of DeepDive (circles); the DeepDive secondary sources, which are quiescent galaxies selected as quiescent galaxies in archival search in Section \ref{sec:archive} (triangles); and quiescent galaxies from the archive (squares). Their color corresponds to their redshift. As in the left panel, if the SFRs from the SED fitting are below $1\, M_\odot {\rm yr^{-1}}$, they are replaced by a $2\sigma$ upper limit. The SFMS at $z=1,\, 2,\, 3,\, {\rm and}\, 4$ of \citet{schreiber_2015} and 1dex below of them are shown in solid and dashed lines, respectively. We note that the SFR shown here is based on the SED fitting with DJA photometry for both the DeepDive sample and quiescent galaxies in DJA to make the method of measuring the photometry consistent.}
    \label{fig:SFR}
\end{figure*}

The obtained narrow \ha\ fluxes were converted to the SFR following \citet{kennicutt-evans_2012}. We corrected the dust extinction using the dust attenuation of the stellar light from the SED fitting, though the dust attenuation is relatively low ($\sim0.1-0.6$, see Table \ref{tab:maintargets}) and has a large uncertainty. This is based on the assumption that the stellar dust extinction is the same as the dust extinction for nebular lines \citep[but see, e.g., ][and many others]{calzetti_2000,kashino_2013}. We note that we could not derive a robust estimate of the dust attenuation using the Balmer decrement, considering that \hb\ is not detected in the spectra of most of our targets. We derive SFRs between 0 and $5$ \myr\ for all galaxies. These values are broadly consistent with the estimates of SFR averaged over 100 Myr from the SED modeling (left panel of Figure \ref{fig:SFR}). Considering their high stellar masses (Table \ref{tab:maintargets}), such low levels of SFRs confirm the quiescent nature of our targets. Moreover, this estimate is based on the strong assumption that the \ha\ emission is solely from star formation, which, if untrue, could bias the SFR toward high values \citep{Jespersen2025_SED_optical_IR}. Considering the possible presence of AGN ionization as commonly found in distant quiescent galaxies \citep[e.g.,][]{kriek_2009,marsan_2017,belli_2019,carnall_2023b,davies_2024,park_2024,baker_2024,kokorev_2024,deGraaff_2024,barrufet_2024,ito_2025a,bugiani_2025}, and confirmed in a handful of our targets, this estimate of the instantaneous SFR should be considered as an upper limit.  

\section{Archival search for grating spectra of quiescent galaxies up to $z\sim5$}\label{sec:archive}
To put our DeepDive program in a broader context and systematically trace the redshift evolution of the properties of massive quenched systems, we compiled all publicly available JWST/NIRSpec grating spectra of galaxies with suppressed star formation. These were selected from the 67,099 spectra available in DJA (as of March 26, 2025), as well as from spectra of filler sources observed as part of the DeepDive program. We assigned different archival spectra to the same source by allowing for a maximum projected separation of $0.2\arcsec$ between the slit positions. Since we aim to study the stellar spectra of quiescent sources in the rest-frame optical, we selected only those sources with grating spectra covering wavelengths shorter than $\lambda_{\rm rest}<0.5\, {\rm \mu m}$, to focus on spectra comparable to those of the main DeepDive targets. We further imposed a minimum threshold on the median signal-to-noise ratio per pixel at $0.4\, {\rm \mu m}<\lambda_{\rm rest}<0.5 \, {\rm \mu m}$ greater than five to ensure that the stellar continuum emission was reliably detected. 

Next, we compiled the available HST and JWST/NIRCam photometry for each target in DJA. We will henceforth focus on sources whose photometry data are in the DJA photometric catalog. Aperture-corrected total photometry measured within $0\farcs5$ apertures is adopted, as derived for DeepDive main targets with {\sc SEP} using the original images without PSF matching. Existing photometry is also included for sources observed in the same MSA masks as the main DeepDive targets DD-ID 78, 96, 115, 170, and 179. Using this photometry, flux corrections were applied to the spectra in the same way as for the primary DeepDive targets. Stellar population synthesis modeling of the photometry for all archival sources was also performed using the setup described in Section \ref{sec:SEDfitting}.

We then classified galaxies in this continuum-selected sample as ``quiescent'' if they satisfy one of the following three traditional yet distinct criteria: 
\begin{enumerate}
    \item \dn\ index selection. We selected sources with $D_n 4000>1.2$ as quiescent galaxies. This index was derived from the spectra following the definition in \citet{balogh_1999}. The threshold roughly corresponds to the strength of the break of a simple stellar population with the age of a few $100\, {\rm Myr}$ with solar-metallicity and no dust extinction \citep[e.g.,][]{d'eugenio_2021,wu_2021}. To ensure the quality of the \dn\ measurement, we required its uncertainty to be less than 20\% of the value.
    \item $UVJ$ color selection. We used the following selection criteria for $1.0<z<2.0$ from \citet{williams_2009}, but allowed the extra 0.2 mag margin to select sources outside of the selection criteria to account for the uncertainties on the rest-frame colors, following \citet{valentino_2023}:
        \begin{align*}
        (U-V) &> 1.1,\\
        (V-J) &< 1.8,\\ 
        (U-V) &> 0.88\times(V-J)+0.29. 
        \end{align*}
    \item Specific star formation rate (sSFR) selection. Sources were selected as quiescent galaxies if they have an sSFR from the purely photometric SED fitting more than one dex below the star formation main sequence (SFMS) of \citet{schreiber_2015} at their redshift and stellar mass. The choice of the SFMS has a negligible impact on the selection. For instance, adopting the \citet{popesso_2023} definition, all quiescent galaxies selected using the \citet{schreiber_2015} SFMS are retained, except for a single source. Similarly, even if we apply the alternative criterion which \citet{carnall_2020, Carnall_2022qg} applied, i.e., ${\rm SFR}<0.2/(t_{\rm age}/{\rm yr})$, where $t_{\rm age}$ is the age of the Universe, yields the same result as the selection based on the SFMS of \citet{popesso_2023}.
\end{enumerate}

Finally, we visually inspected the spectra of all selected quiescent candidates to exclude obvious contaminants and flag less certain quiescent galaxies. Some spectra clearly exhibit significantly broad or strong emission lines without pronounced Balmer breaks, contaminating the rest-frame optical bands and leading to misclassifications (see Figure \ref{fig:archive-susspectra}). Eight such objects were removed, resulting in a contamination fraction of $\sim7\%$. 
We conservatively removed the other nine sources that have specific star formation from the photometric SED modeling within $2\sigma$ \citep[i.e., 0.4\, dex, e.g., ][]{speagle_2014} of the SFMS at their redshift. These objects were originally selected only based on their \dn. Seven out of these objects have blue colors in their outer regions, suggesting ongoing star formation activity in extended disks, while the spectra only trace the central red, passive bulges. Such spatial decoupling in the distribution of the star formation activity is the reason why we simultaneously measure high SFRs from the integrated photometry and a high \dn\ typical of old stellar populations imprinted in the spectra mapping the central cores. Lastly, one galaxy -- host to a Type Ia supernova and gravitationally lensed \citep{frye_2024} -- was also excluded from the sample.

The final sample of spectroscopically confirmed galaxies identified as quiescent based on at least one of the criteria mentioned above consists of $126$ unique galaxies at $1<z<5$. 81, 110, and 118 sources are selected as quiescent galaxies with the \dn\ index, $UVJ$ colors, and sSFR selection, respectively. Eighteen archival sources are selected as quiescent from the DeepDive survey by one of the above three criteria. They are hereafter referred to as the ''secondary" DeepDive targets. Thirty-three sources lack the full wavelength range needed for measuring the \dn. All of the selected spectra are obtained by medium-resolution spectroscopy. Appendix \ref{app:archivallist} presents the full list of programs from which the archival spectra were obtained. In the released catalog, columns indicate which selection criteria each QG satisfies (${\rm QG_{D_n 4000}}$, ${\rm QG_{UVJ}}$, ${\rm QG_{sSFR}}$). These flags enable the construction of QG samples based on specific selection methods of interest.

\begin{figure}
    \centering
    \includegraphics[width=0.8\linewidth]{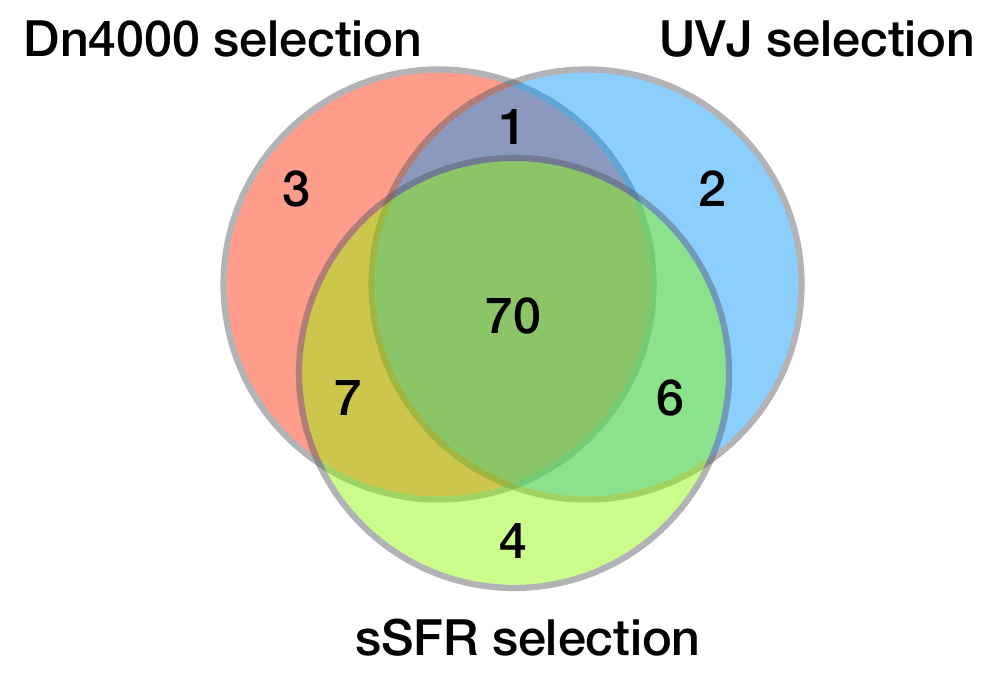}
    \caption{Venn diagram of the number of selected quiescent galaxies in Section \ref{sec:archive} based on \dn\ index, $UVJ$, and sSFR criteria, respectively. Here, we focus on 96 quiescent galaxies with a \dn\ measurement.}
    \label{fig:Ven_archive}
\end{figure}

This sample enables a comparison of the three different criteria for selecting quiescent galaxies. Among 93 archival quiescent galaxies with both \dn\ measurements, 70 sources ($\sim75\%$) satisfy all three criteria. The overlap between the selections is summarized in Figure \ref{fig:Ven_archive}. The $UVJ$ and sSFR samples both significantly and equally overlap with the \dn\ sample, with $90$\% of each included within it. 87\% of the sSFR sample is also included in the $UVJ$ sample.

We emphasize that the spectroscopic and photometric datasets of this large sample of 136 quiescent galaxies, combined with DeepDive main targets, were homogeneously reduced and analyzed, allowing for a systematic comparison of their constituents. For display purposes, we show a subsample of the archival spectra in Figure \ref{fig:archive-spectra}. All spectra analyzed herein, including those of DeepDive main targets, are publicly available online\footnote{\url{https://doi.org/10.5281/zenodo.18508154}}. Those of the removed sources as contaminants are also stored, which can be interesting to study physical processes such as inside-out quenching. Table \ref{tab:archivalQG}, also available online, summarizes their properties, which will be further discussed in the next Section. 

\section{Basic characteristics of quiescent galaxies with NIRSpec medium resolution spectra}
\label{sec:properties}

We present the basic properties of the DeepDive main targets in this Section. By comparing quiescent galaxies systematically selected in Section~\ref{sec:archive}, we put them into a general context. We note that more detailed analyses, summarized in Section~\ref{sec:sciencegoal}, will be presented in our upcoming papers.

\subsection{Stellar masses, star formation rates, and redshifts}\label{sec:mstarsfr}
\begin{figure}
    \centering
    \includegraphics[width=0.95\linewidth]{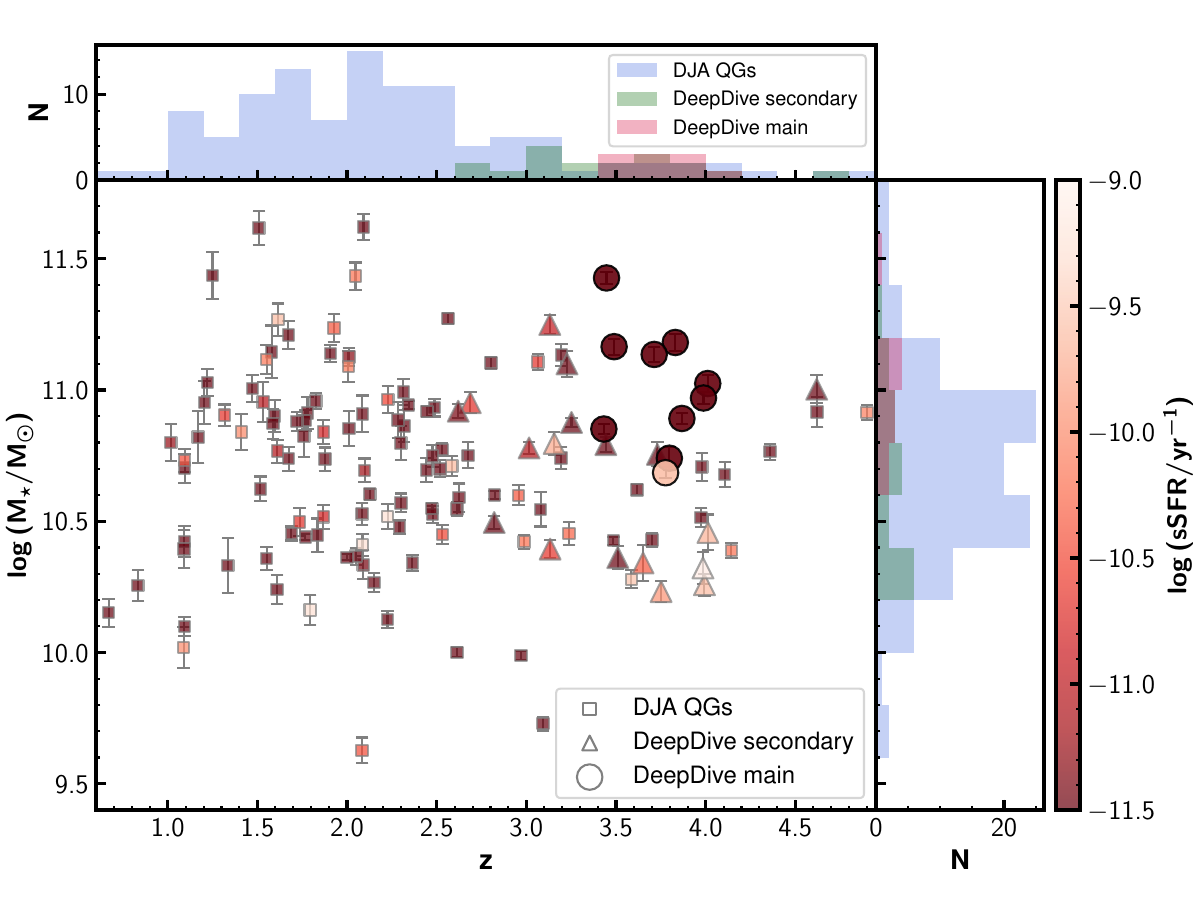}
    \caption{Stellar mass-redshift plane occupied by the main targets of the DeepDive program (large filled circles), the DeepDive secondary sources selected as quiescent galaxies (triangles), and quiescent galaxies from the archive (squares). Sources without photometry are not displayed. The top and right panels display the stellar mass and redshift distributions of each sample: the main and secondary DeepDive targets in red and green, and the rest of the archival quiescent galaxies in blue.}
    \label{fig:Mstar-redshift}
\end{figure}

Figure \ref{fig:Mstar-redshift} shows the distribution of our DeepDive and archival quiescent galaxy samples in the stellar mass and redshift plane. The redshift distribution of the archival sources peaks at $z\sim2$ and their stellar masses range between $\log{(M_\star/M_\odot)}=10-11$. On the other hand, the DeepDive main targets are located at a higher redshift ($z=3.4-4$) by selection. Moreover, at fixed redshift, DeepDive quiescent galaxies have higher stellar mass ($\log{(M_\star/M_\odot)}=10.8-11.5$) than archival sources. This highlights the strong complementarity between DeepDive and the archival observations. The right panel of Figure \ref{fig:SFR} shows the stellar mass-SFR relation, where both quantities were derived from the photometric SED modeling (Section \ref{sec:SEDfitting}). Most of the selected sources are well below the SFMS of \citet{schreiber_2015}.

\begin{figure}
    \centering
    \includegraphics[width=0.8\linewidth]{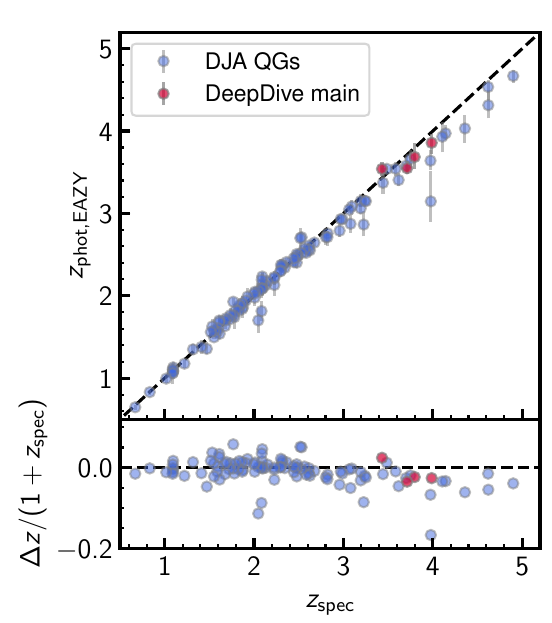}
    \caption{Comparison between the spectroscopic redshift obtained from NIRSpec grating spectra of quiescent galaxies and the photometric estimate from {\sc EAZY-py} taken from DJA. Red markers indicate the locations of the main DeepDive targets that fall within known cosmological JWST fields covered by public catalogs in DJA, while blue symbols denote sources identified through the archival search. The bottom panel shows the deviation from the one-to-one relation.}
    \label{fig:specz-photz}
\end{figure}

This large sample of reliable spectroscopic redshifts enables us to evaluate the quality of photometric redshifts for the population of distant quiescent galaxies. Figure \ref{fig:specz-photz} shows the relation between the spectroscopic and photometric redshifts listed in public DJA photometric catalogs and derived with {\sc Eazy-py}. For a fair comparison, we limited the analysis to the sources in cosmological fields with ample JWST and HST photometric coverage. The photometric redshifts agree well with the spectroscopic estimates, especially at $z < 3$. The normalized median absolute deviation for the whole sample, defined as $\sigma_{\rm NMAD} = 1.48\times{\rm median}\left(|(z_{\rm phot}-z_{\rm spec})-{\rm median}(z_{\rm phot}-z_{\rm spec})|/(1+z_{\rm spec})\right)$, is $\sigma_{\rm NMAD} = 0.022$, and the bias, defined as $b={\rm median} (z_{\rm phot}-z_{\rm spec})$, is $b=-0.01$. However, the bias becomes more significant ($b=-0.13$) if only sources at $z=3-5$ are used for calculation, whereas the $\sigma_{\rm NMAD}$ does not change significantly $\sigma_{\rm NMAD}=0.027$). This indicates that photometric redshifts of the most distant quiescent galaxies are systematically underestimated.

\subsection{X-ray detections}
We searched for possible X-ray detections of our total sample of 136 quiescent galaxies in the public Chandra X-ray catalogs of the COSMOS \citep{civano_2016}, UDS \citep{Kocevski_2018}, AEGIS \citep{Nandra_2015}, GOODS-S \citep{Luo_2017}, and GOODS-N \citep{Alexander_2003} fields. By cross-matching with a maximum separation of $2''$, we identified 17 quiescent galaxies listed in these catalogs (including the two DeepDive main targets mentioned in Section~\ref{sec:deepdive}), corresponding to 13\% of the 127 sources within their coverage. Their rest-frame $2-10$ keV luminosities span $5\times10^{41}-3\times10^{44}\,{\rm erg\,s^{-1}}$, derived from the full-band flux assuming a photon index of 1.8 and neglecting X-ray obscuration. These luminosities are generally higher than those expected from X-ray binaries \citep[$\sim10^{41}\,{\rm erg\,s^{-1}}$; e.g.,][]{Lehmer_2016, Fornasini_2018}. Although the intrinsic detection fraction is uncertain due to varying X-ray depths, this suggests that a non-negligible fraction of the quiescent galaxy sample hosts X-ray–bright AGNs. The X-ray luminosities are also reported in the released catalogs.

\subsection{$UVJ$ color diagram}\label{sec:UVJ}
\begin{figure*}
    \centering
    \includegraphics[width=0.95\linewidth]{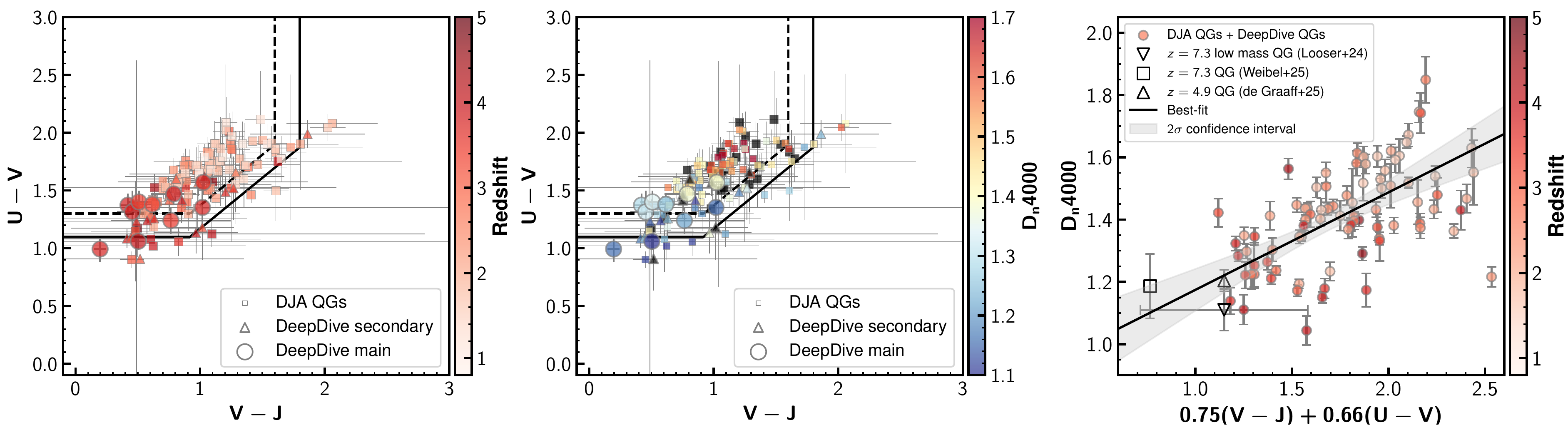}
    \caption{$UVJ$ color properties. Left and middle panels show the $UVJ$ rest-color diagram, where markers are colored by their redshift and \dn, respectively. The meaning of the markers is the same as in Figure \ref{fig:Mstar-redshift}. Sources without \dn\ measurement are colored black in the middle panel. The dashed and solid lines correspond to the threshold in \citet{williams_2009} and that applied in this study, respectively. The right panel shows the correlation between the \dn\ index and the position along the diagonal direction of the quiescent selection box in \citet{williams_2009}, parameterized as $S_Q=0.75(V-J)+0.66(U-V)$. The markers show all $UVJ$-selected quiescent galaxies with \dn\ measurements from grating spectra, colored by their redshift. The black line is the best-fit linear model, and the gray hatched region corresponds to its $2\sigma$ confidence interval. For reference, the location of two recently confirmed massive quiescent galaxies at $z=7.3$ \citep{weibel_2025} and $z=4.9$ \citep{deGraaff_2024}, and a low-mass system at $z=7.3$ \citep{looser_2024} are shown as a black open square, an upward triangle, and a downward triangle, respectively.}
    \label{fig:UVJ}
\end{figure*}

In the left and middle panels of Figure \ref{fig:UVJ}, we show the location of the main DeepDive targets relative to the archival sample in the rest-frame $UVJ$ diagram. As detailed in Section~\ref{sec:archive}, we selected quiescent galaxies, allowing for 0.2dex relaxed constraints compared to the limits in \citet{williams_2009}, and we accounted for the uncertainties in the rest-frame colors. Some sources, however, are even outside of these relaxed limits, but they are classified as quiescent galaxies based on their low sSFR or \dn\ index. Specifically, the ones in the lower-left region outside the selection box (with $U-V\sim1$ and $V-J<1$) are primarily selected based on the sSFR criterion. These sources are likely more recently quenched galaxies and could be recovered using revised selection thresholds proposed in recent studies \citep[e.g.,][]{belli_2019, baker_2024,baker_2025b}.

The left and middle panels of Figure \ref{fig:UVJ} offer different insights on the $UVJ$ selection technique over the covered redshift range ($1<z<5$). Higher redshift sources, including the DeepDive targets, cluster toward bluer colors and lower \dn\ index: recently quenched systems are more easily found at high redshift, often displaying features typical of post-starburst galaxies and benefiting from lower mass-to-light ratios during selection  \citep{forrest_2020b,d'eugenio_2021,antwi-danso_2023}. Moreover, galaxies at higher redshift are expected to be younger on average, given the younger age of the Universe itself.  Although rare, older objects at $z>3$ could be key to deciphering the early formation of massive galaxies and may pose challenges to standard models \citep{steinhardt_2021, Lovell2023_EVS, glazebrook_2024,carnall_2024, jespersen2025_extreme}. The age gradient along the diagonal line of the selection box (here intended as time since the quenching epoch) has been highlighted in several previous works \citep[e.g.,][]{whitaker_2012,whitaker_2013,belli_2019}. This is reflected in the distribution of observed higher \dn\ indices in the upper right region in the central panel in Figure \ref{fig:UVJ}, although the location in the color diagram is also correlated with dust reddening.

We further explore this trend by modeling the observed \dn\ index as a function of the location in the $UVJ$ diagram. To simplify the interpretation of such a trend, we apply a change in coordinates as in \cite{belli_2019} (see also \citealt{fang_2018}). The location perpendicular to the selection box ($(U-V) > 0.88\times(V-J)+0.29$) is parameterized as $S_Q = 0.75(V-J)+0.66(U-V)$. The right panel of Figure \ref{fig:UVJ} shows the relation between the \dn\ index and $S_Q$. Here, only the $UVJ$-selected quiescent galaxies with the \dn\ measurements are shown. The Spearman correlation test using them indicates a moderate correlation between \dn\ and $S_Q$ at a significant level ($r=0.57$, p-value $<$ 0.01). A simple linear model fitting returns
\begin{equation}
    D_n4000 = (0.31\pm0.05) S_Q+(0.86 \pm 0.08),
    \label{eq:dn4000_uvj}
\end{equation}
with an observed scatter of $10\%$. Sources with $S_Q>2.3$ were excluded from the fitting procedure, as potential dusty contaminants with a median $A_V\sim1.3$\, mag. The moderate scatter is likely due to the variation in dust extinction across the sample. \citet{Cheng_2025} also reports a scatter, which conducts a comparable analysis of the stellar age of galaxies in the $UVJ$ diagram. The estimates of \dn\ index and the location of the $UVJ$ diagram are almost independent here, with the only connection established during the calibration of the spectra using photometry (Section~\ref{sec:fluxcorrection}).

The trend, interpreted here as due to the stellar age, is also a practical tool to map classical $UVJ$ color selections of quenched galaxy candidates, especially in use at lower redshifts, with the strength of the break gaining renewed popularity in the high redshift community \citep[e.g.,][]{weibel_2025}. For reference, we show the location of the $z=4.9$ and $z=7.3$ massive quenched objects in \citet{deGraaff_2024} and \citet{weibel_2025}, respectively, and the lower-mass system in \citet{looser_2024} at $z=7.3$, probably in a transitory phase of suppressed star formation. They overall follow the obtained relation between the \dn\ and location of the $UVJ$ color, even though they were not used in determining the relation. The massive galaxies at $z=4.9$ and $z=7.3$ fall within the $2\sigma$ confidence interval, whereas the low-mass galaxy at $z=7.3$ has a marginally lower \dn\ than that expected from its location in the $UVJ$ plane, which might be due to the large uncertainty in its rest-frame colors. This suggests that the obtained scaling relation is valid for a wide stellar mass range and at higher redshifts.

\subsection{Composite spectra as a function of \dn}\label{sec:stack}
In the previous sections, we introduced our DeepDive program and the archival search, and we assembled a large sample of quiescent galaxies with high-quality grating spectra from JWST/NIRSpec. By construction, each spectrum is characterized by high S/N on the stellar continuum, which will allow us to study their physical parameters (Section~\ref{sec:sciencegoal}). Here, we showcase what can be achieved when we push even further the capabilities of the instrument and stack tens of spectra to derive the average spectral shapes of massive quiescent galaxies at different times after their main quenching epoch. 

\begin{figure*}
    \centering
    \includegraphics[width=0.95\linewidth]{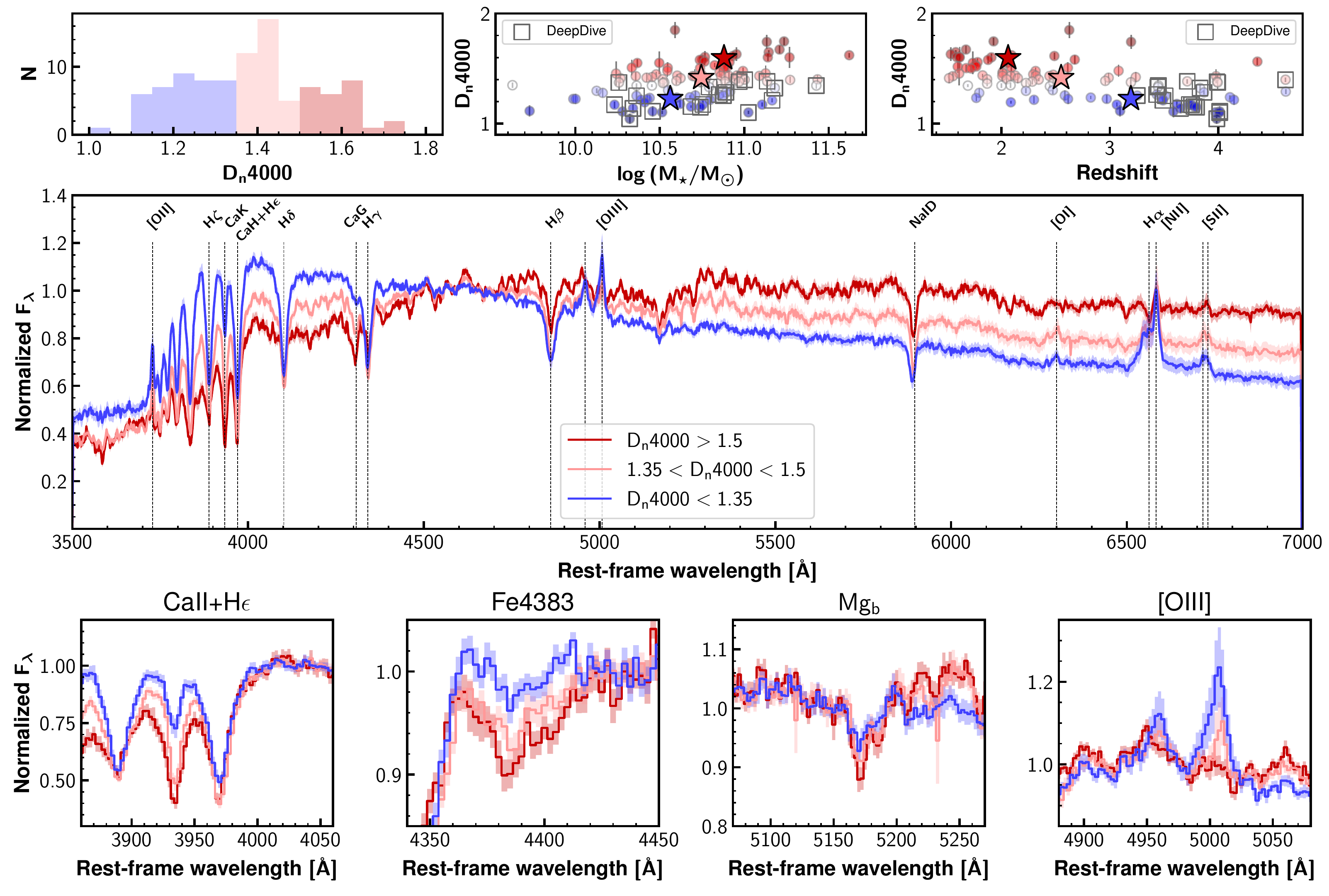}
    \caption{Top row, from left to right: Distribution of the \dn\ index, \dn\--stellar mass, and \dn\--redshift planes. Three subsamples have been constructed and color coded based on \dn. The stars in the middle and right panels mark the median values of each \dn\ subsample. Sources from DeepDive are highlighted with open black squares. Middle row: Average-stacked spectra of the three \dn\ subsamples. The same color is used as in the top row. The spectra are normalized by the median flux at $\lambda_{\rm rest}=0.44-0.48\, {\rm \mu m}$. Bottom row: Zoom-in view of the average stacked spectra around CaII+H$\epsilon$, Fe4383, ${\rm Mg_b}$, and the \oiii\ lines.}
    \label{fig:stackspec-Dn4000}
\end{figure*}

We divided our combined sample based on their \dn\ index values. We thus removed sources whose spectra do not cover the $0.385\, {\rm \mu m}<\lambda_{\rm rest}<0.45\, {\rm \mu m}$ range, retaining a final sample of 98 sources. We bin these galaxies into a high ($D_n 4000 > 1.5$), an intermediate ($1.35 < D_n 4000 < 1.5$), and a low ($D_n 4000 < 1.35$) \dn\ samples. These bins contain 25, 34, and 39 sources, respectively. We show in Figure \ref{fig:stackspec-Dn4000} the distributions of the \dn\ index, stellar mass, and redshifts. The low \dn\ sample primarily consists of sources at higher redshift (with a median of $z\sim3$) than the other samples. 16 out of 39 ($41\%$) spectra of the low \dn\ sample were obtained as part of DeepDive, highlighting again how this program complements other efforts to characterize massive, quenched systems in the distant Universe by covering relatively recently quenched systems. The high \dn\ sample consists of more massive galaxies, partially because they are at a lower redshift. Nonetheless, the difference in stellar masses among the three subsamples is marginal (within 0.5 dex). We also note that if we used quiescent galaxies based on a single selection criterion (e.g., low sSFR), the following results would not change significantly.

To create the average spectra, we followed the method presented in \citet{onodera_2012}. The spectra were first de-redshifted and normalized by the average flux at $0.44\, {\rm \mu m}<\lambda_{\rm rest}<0.48\, {\rm \mu m}$. They were then resampled on a 2.5\AA\ rest-frame wavelength grid, which is similar to the sampling size of NIRSpec medium resolution spectra, by linearly interpolating the original spectra. We then computed the average spectra by optimally weighting each spectrum by the inverse square of its noise array. The uncertainties ($\sigma$) on the average spectra were estimated by applying the jackknife method: 
\begin{equation}
    \sigma^2 = \frac{N-1}{N} \sum_{i=1}^{N} (f-f_i)^2,
\end{equation}
where $N$ is the number of spectra used, $f$ is the average spectrum determined by using all $N$ spectra, and $f_i$ is the average spectrum determined by using the $N-1$ spectra removing $i$-th spectrum. The possible bias due to the limited sample size is corrected by the following equation:
\begin{equation}
    f' = f- (N-1)\left(\langle f_i\rangle-f\right),
\end{equation}
where $f'$ is the bias-corrected average spectrum and $\langle f_i\rangle$ is the mean of the $f_i$ spectra. We use the $f'$ as the average spectrum of each subsample.

Figure \ref{fig:stackspec-Dn4000} shows the stacked spectra of the galaxies grouped according to their \dn\ index. Thanks to the large number of objects in each subsample, a high signal-to-noise ratio is achieved for the stacked spectra. The median signal-to-noise ratio per pixel at $0.4\, {\rm \mu m}<\lambda_{\rm rest}<0.5\, {\rm \mu m}$ of the stacked spectra of low-\dn, mid-\dn, and high-\dn\ sample is ${\rm S/N} = 63$, $73$, and $69$, respectively. The spectra clearly become progressively redder with increasing \dn\ index, reflecting the older stellar populations in higher \dn\ samples. Additionally, the absorption line strengths differ among the three stacked spectra. In particular, the CaII absorption lines are stronger in the higher-\dn\ samples, consistent with their older stellar populations. Conversely, the brightest \oii, \oiii, and \ha\ emission lines are found in the lowest-\dn\ bin. One possible explanation is AGN activity, as recently quenched galaxies with low \dn\ values may have a higher likelihood of hosting AGNs -- similar to what is observed in post-starburst quiescent galaxies at high redshift \citep[e.g.,][]{belli_2019,deGraaff_2024,bugiani_2025,nanayakkara_2025}. Alternatively, the \oiii\ emission could originate from residual star formation, suggesting that galaxies in the lower-\dn\ bin retain higher levels of ongoing star formation.

\begin{figure}
    \centering
    \includegraphics[width=0.9\linewidth]{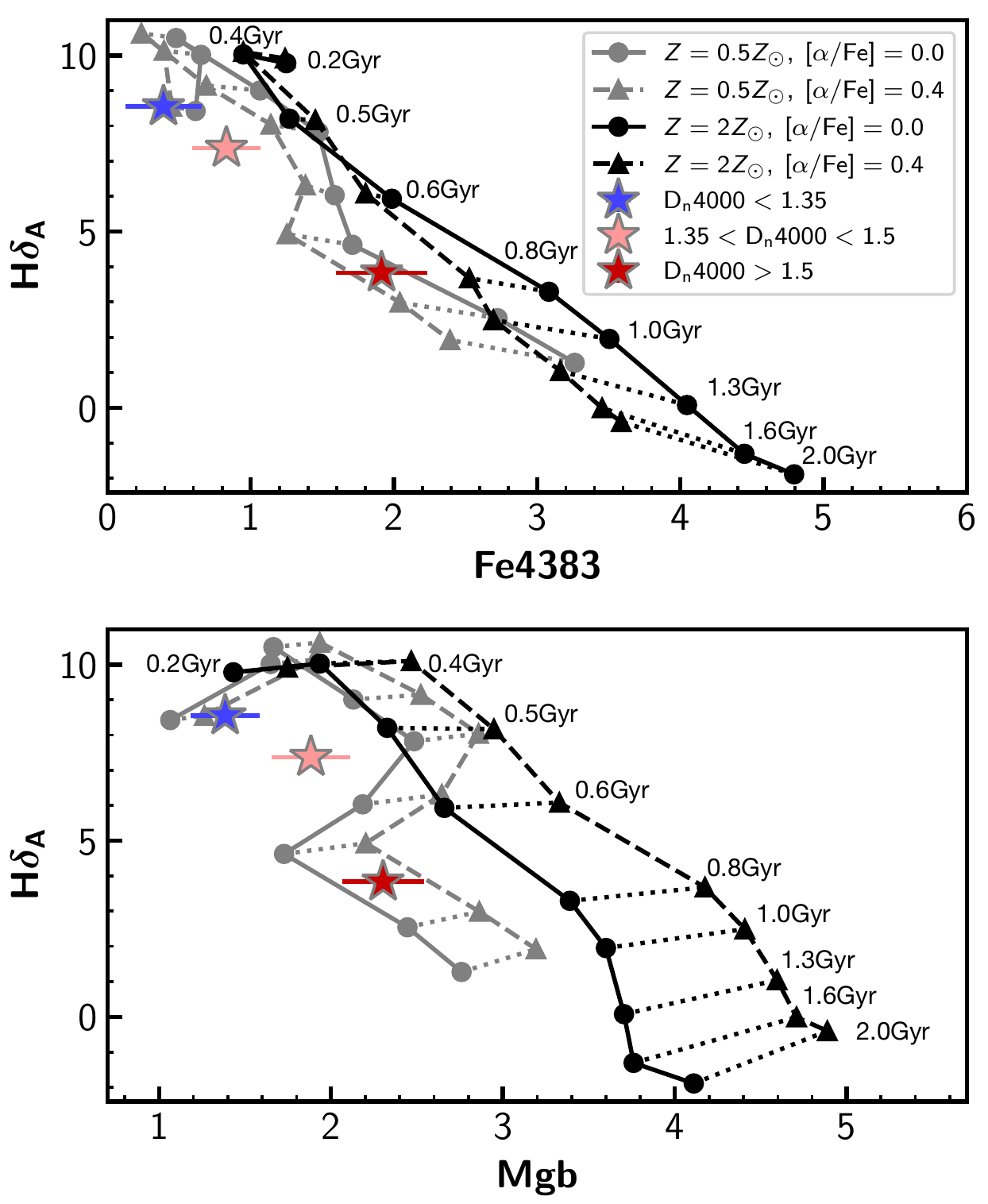}
    \caption{Top panel: H$\delta_A$ vs. Fe4383 for the three \dn-subsamples, shown as filled stars colored as Figure \ref{fig:stackspec-Dn4000}. Black and gray lines represent the predicted index strengths from the BPASS model with stellar metallicities $Z = 2\, Z_\odot$ and $Z = 0.5\, Z_\odot$, respectively. The solid (dashed) line is computed for $[\alpha/{\rm Fe}] = 0.0\, (0.4)$. The age values corresponding to each data point of the $Z = 2\, Z_\odot$ case are indicated. Bottom panel: H$\delta_A$ vs. ${\rm Mg_b}$. The color coding and the lines have the same meaning as those in the top panel.}
    \label{fig:Lick}
\end{figure}

Lastly, several faint metal absorption lines are significantly detected even in the average spectrum of the lowest-\dn\ sample at a median redshift of $z = 3$. Figure~\ref{fig:Lick} presents the measurements of H$\delta_A$, Fe4383, and ${\rm Mg}_b$, following the Lick index definitions \citep{Trager_1998}, as representative age- and metallicity-sensitive spectral features. These indices are well constrained and show clear variations as a function of \dn. We estimate Fe4383 (${\rm Mg}_b$) indices of $1.91\pm0.32$\AA, $0.83\pm0.24$\AA, and $0.39\pm0.27$\AA\ ($2.30\pm0.24$\AA, $1.88\pm0.22$\AA, and $1.38\pm0.20$\AA) for the high-\dn, mid-\dn, and low-\dn\ samples, respectively. In Figure~\ref{fig:Lick}, we also show the predicted index strengths from BPASS single stellar population models \citep[v2.3,][]{Byrne_2022} with different metallicities, $\alpha$-enhancements, and ages. They are obtained after smoothing the templates to match their resolution to that of the NIRSpec medium grating spectra with a stellar velocity dispersion of $250\, {\rm km\, s^{-1}}$. Although we here adopt simplified assumptions of single stellar populations and closed-box evolution for illustrative purposes, the comparison with models demonstrates that both stellar age and metallicity can be meaningfully constrained. The obtained average index values of three subsamples follow the age track, as expected from their \dn\ values, at similar metallicities. The detection of these faint features, together with the quality of the line-strength measurements, highlights the potential for systematic studies of metal enrichment histories in massive galaxies in the early Universe (see Section~\ref{sec:sfh-metal}).

We note that the stacked spectra presented here are likely dependent on other properties, such as stellar mass or the shape of the star formation history. The detailed correlation with spectra and other parameters will be investigated based on individual source searches, as stacking with a smaller number of sources (e.g., fewer than ten sources) might be affected by low number statistics.

%%%%%%%%%%%%%%%%%%%%%%%%%%%%%%%%%%%%%%%%%%%%%%%%%%%%%%%%%%%%%%
\section{Science objectives of DeepDive}
\label{sec:sciencegoal}
The purpose of this work is to introduce our program and to demonstrate how the DeepDive sample compares with a homogeneously selected set of similarly quenched galaxies, all benefiting from comparable data quality and coverage. While detailed analyses of the formation and evolution of massive quenched galaxies will be presented in forthcoming studies, we take this opportunity to briefly outline the main science cases that the DeepDive spectra will inform. Complemented by the large sample of archival spectra of quiescent galaxies compiled in this paper, population studies will now be possible. We stress that all of the science cases are uniquely enabled by the grating spectra ($R\geq 1000$).

\subsection{Star formation and metal enrichment histories}\label{sec:sfh-metal}
Recent JWST studies of massive quiescent galaxies suggest that their formation and quenching timescales are short, consistent with forming very early in the Universe and quenching rapidly via abrupt feedback mechanisms \citep[see e.g.,][]{carnall_2023b,carnall_2024, Slob_2024, park_2024,deGraaff_2024}. These formation timescales are encapsulated in the chemical abundance patterns of galaxies through powerful diagnostics such as $\alpha$-enhancement [$\alpha$/Fe] \citep[e.g.,][]{kriek_2019, jafariyazani_2024, beverage_2021, beverage_2024, beverage_2025}, due to the short timescales on which $\alpha$ elements (e.g., Mg, O, Ne, Si) are produced in Type II (core-collapse) supernovae relative to iron-peak elements, which are produced mainly in Type Ia supernovae \citep[see e.g.,][]{woosley_weaver_1995,thomas_2003}. Owing to the high-quality spectroscopy afforded by the DeepDive program, we will be able to extract robust star-formation histories and metal abundances (e.g., [Mg/Fe]) at $z>3$ to directly constrain these evolutionary timescales for statistically significant samples \citep{Hamadouche_2026}.

\subsection{Stellar dynamics}
The high S/N detection of stellar absorption features allows us to derive the stellar velocity dispersion. This parameter is known to be correlated with various fundamental properties of galaxies in the local Universe, as seen in the fundamental plane \citep{djorgovski_1987}. The stellar velocity dispersion has been measured in a handful of quiescent galaxies at $z\sim3-4$ using ground-based telescopes \citep[e.g.,][]{tanaka_2019,saracco_2020, esdaile_2021,forrest_2022}, but the estimates are affected by large uncertainties, and the overall number of measurements is still severely limited. The DeepDive program will double the number of measurements of accurate stellar velocity dispersions at these redshifts. While still affected by uncertainties due to the geometry of the systems and possibly unresolved stellar rotation in compact objects \citep[e.g.,][]{Slob_2025}, the estimate of the stellar velocity dispersion will allow for study of the existence of a fundamental plane of massive quenched systems up to $1.5$ Gyr after the Big Bang and to trace its time evolution with the rest of the archival sample \citep[e.g.,][]{mendel_2020, stockmann_2021, vanderwel_2022}. Thanks to NIRCam imaging, from which we can determine the half-light radius (e.g., \citealt{ito_2024, Kawinwanichakij_2025, Yang_2025}, Scarpe et al. in prep. for the size measurement of quiescent galaxies), we will also be in the position to estimate dynamical masses and put a constraint on the IMF by comparing it with stellar masses from SED modeling (Ito et al. in prep.). 

\subsection{Ionized interstellar medium properties} 
As visible in both the individual and average spectra presented above, we detect significant emission lines from the ionized interstellar medium in several massive quenched galaxies. The presence of such emission lines, powered by AGN, residual SF, shocks, or a combination of these mechanisms, is now a rather well-established finding allowed by JWST observations \citep{carnall_2023b,d'eugenio_2023,baker_2024,park_2024,bugiani_2025}. Interestingly, the stacked spectra in bins of \dn\ seem to suggest the presence of stronger emission in more recently quenched systems (Figure \ref{fig:stackspec-Dn4000}), and the line ratios might point to a significant contribution of AGN activity to the powering of the emission. The large available sample compiled in this work will allow for the exploration of trends on a solid statistical basis (Kakimoto et al. in prep.). 

\subsection{Outflows}
\label{sec:outflow}
Recent observations indicate that neutral gas outflows are common in quenched galaxies beyond the local universe \citep{davies_2024,belli_2024, sunExtremeNeutralOutflow2025}. Gas outflows could be powered by AGN, supernovae, and stellar winds from young stars. They play a crucial role in the baryon cycle of galaxies, regulating their evolution by removing gas and redistributing energy and momentum into the surrounding medium. Half of the DeepDive main targets (and more in the secondary targets) show signs of blue-shifted Na \textsc{i} D$\lambda\lambda$5891, 5897 absorption, a standard tracer of such outflows. The DeepDive sample will unlock the study of neutral gas outflows in uncharted redshift and mass regimes (Figure \ref{fig:Mstar-redshift}), the missing transition phase between ancient star-forming galaxies and present-day quiescent galaxies \citep{Zhu_2026}.

\section{Summary}
We presented an overview of the DeepDive program, which aims to characterize the physical mechanisms driving the formation and evolution of early massive quiescent galaxies using medium-resolution spectra obtained with JWST/NIRSpec, along with an archival search for additional grating spectra of quiescent galaxies. In this work, we discussed the following points:

\begin{itemize}
    \item We introduced the primary DeepDive sample of ten color-selected quenched systems at $z\sim3-4$, whose high stellar masses ($\mathrm{log}(M_\star/M_\odot)>10.7$) and young stellar ages fill a portion of the parameter space so far scarcely explored with deep JWST spectra.
    \item We focused on the presentation of newly acquired four-band NIRCam images and G235M/F170LP NIRSpec spectra as part of DeepDive, and we delved into a few technical details of the data reduction. In particular, we concentrated on a careful flux calibration of the newly extended wavelength ranges covered by v4 spectra from the latest version of the {\sc MSAExp} pipeline, developed as part of the DJA initiative. The wavelength extension results are crucial to cover spectral features such as the \ha+\nii\ complex in all DeepDive primary targets.
    \item Based on measurements of the \ha\ emission, inaccessible from the ground for this redshift, we confirmed the quiescence of our main targets ($\mathrm{SFR}\sim0-5$ \myr), and highlighted the presence of AGNs for two sources as revealed by broad wings in the emission lines.
    \item To place the DeepDive targets in a broader context, we carried out an extensive search for archival quiescent galaxies with data quality and coverage comparable to those of the main DeepDive sample. Using three classical selection methods -- the strength of the 4000\AA\ break (quantified via the \dn\ index), the $UVJ$ color diagram, and a specific star formation rate (sSFR) threshold -- we identified $126$ unique sources with high-S/N NIRSpec grating spectra. The redshift distribution of the archival quiescent galaxies peaks at $z\sim2$. Combined with the DeepDive targets, the resulting sample spans a redshift range of $1<z<5$ and covers 1.5 dex in stellar mass. All spectra were reduced and calibrated using the same methodology as for the DeepDive primary targets, allowing for consistent and systematic comparisons.
    \item We found an approximately $90$\% overlap between these three selection criteria for quiescent galaxies, while $75$\% of the spectroscopic sample simultaneously satisfies all selections.
    \item We mapped the \dn\ indices measured from the spectra onto $UVJ$ rest-frame colors from the photometric modeling. As a natural consequence of the shorter time available to assemble and quench and selection criteria privileging bright sources, high-redshift quenched systems are caught closer to their main quenching epoch, thus displaying bluer colors and lower \dn\ index. The parameterization in Eq. \ref{eq:dn4000_uvj} is a useful tool for comparing the classical $UVJ$ selection, commonly used at $z\lesssim2-3$, with the break strength criteria, which is increasingly popular for defining quenched systems at higher redshifts.
    \item Finally, we generated average spectra for massive quiescent galaxies at different post-quenching stages as tracked by \dn. The overall spectra are redder for higher \dn\ samples. Also, absorption line features change along with the \dn\ value, as expected from the correlation between age and \dn\ index. The emission lines are stronger for the lower \dn\ value sample, possibly due to residual star formation or AGN activity in more recently quenched galaxies. The achieved S/N is such that we detect various faint metal absorption features that were previously detectable only at low redshift, showing that we can constrain the metallicity even for quiescent galaxies at $z\geq3$.
\end{itemize}

Such a large statistical sample, encompassing wide intervals of stellar masses, redshifts, ages, and several other properties, will be key to propelling research on the formation and evolution of massive quenched galaxies—following the approach described in Section~\ref{sec:sciencegoal} -- beyond the current ``discovery'' phase and into one of statistical consolidation, where robust statistics will reveal the underlying physics. To help foster new studies, all homogeneously reduced and analyzed data used in this work, both from DeepDive and the archival search, are publicly released with this work\footnote{\url{https://doi.org/10.5281/zenodo.18508154}}.

\begin{acknowledgements}
We appreciate the helpful comments and suggestions from the anonymous referee that improved the manuscript. KI, FV, and PZ acknowledge support from the Independent Research Fund Denmark (DFF) under grant 3120-00043B. This study was supported by JSPS KAKENHI Grant Numbers GB25KJ1331, JP23K13141, and JP25K07361, and in part by The Graduate University for Advanced Studies, SOKENDAI. PFW acknowledges funding through the National Science and Technology Council grants 113-2112-M-002-027-MY2. This work is based on observations made with the NASA/ESA/CSA James Webb Space Telescope. The data were obtained from the Mikulski Archive for Space Telescopes at the Space Telescope Science Institute, which is operated by the Association of Universities for Research in Astronomy, Inc., under NASA contract NAS 5-03127 for JWST. These observations are associated with program \#3567. Some of the data products presented herein were retrieved from the Dawn JWST Archive (DJA). DJA is an initiative of the Cosmic Dawn Center, which is funded by the Danish National Research Foundation under grant DNRF140. 
\end{acknowledgements}

\bibliographystyle{aa} % style aa.bst
\bibliography{bib_deepdive} % your references Yourfile.bib

\begin{appendix}

\section{Summary of the available photometry of the DeepDive main targets}
Table \ref{tab:photoinfo} summarizes the fields where the main DeepDive targets are located, the references to the works that originally reported them, and the list of the available JWST and HST bands covering each field.
\begin{table*}[]
    \small
    \centering
    \caption{Summary of the available photometric information.}
    \begin{tabular}{lcccccc}
    \toprule
    DD-ID & Field\tablefootmark{a} & Name\tablefootmark{b} 
 &Ref.\tablefootmark{c} &$t_{\rm exp, DD}$\, [sec]\tablefootmark{d} &  Available JWST wavelength\tablefootmark{e} & Available HST wavelength\tablefootmark{f}\\
    \midrule
    78 &  SXDS & SXDS-27434 & 1, 2  & 5908.5& 1.5, 2, 3.5, 4.4  &-\\
    80 &  PRIMER-UDS & ZF-UDS-6496 & 3, 4 & - & 0.9, 1.15, 1.5, 2, 2.7, 3.5, 4.1, 4.4 & 0.44, 0.6, 0.8, 1.25, 1.4, 1.6 \\
    96 &  SXDS  & SXDS-10017 & 2 & 11817& 1.5, 2, 3.5, 4.4 & -\\
    106 & COSMOS-Web & COS-DR1-113684 & 5 &2538.5 &1.15, 1.5, 2.7, 4.4 & 0.8, 1.6\\
    111 & PRIMER-UDS & ZF-UDS-3651 & 3, 4 & -  & 0.9, 1.15, 1.5, 2, 2.7, 3.5, 4.1, 4.4 & 0.44, 0.6, 0.8, 1.25, 1.4, 1.6 \\
    115 & COSMOS & COS-466654 & 2 & 10504& 1.5, 2, 3.5, 4.4&-\\
    134 & PRIMER-COSMOS & ZF-COS-20115 & 4, 6, 7 & 10504 & 0.9, 1.15, 1.5, 2, 2.7, 3.5, 4.1, 4.4 & 0.6, 0.8\\
    170 & XMM-LSS & XMM-VID3-2457 & 5 & 3457.6& 1.5, 2, 3.5, 4.4 & 1.6\\
    179 & SXDS & XMM-VID3-2075 &5 &2275.9 & 1.5, 2, 3.5, 4.4&-\\
    196 & CEERS & EGS-31322 & 3, 4 & 8490.7 & 1.15, 1.5, 2, 2.7, 3.5, 4.1, 4.4&0.6, 0.8, 1.25, 1.4, 1.6\\ 
    \midrule
    \bottomrule
    \end{tabular}
    \tablefoot{
    \tablefoottext{a}{Sources listed in the SXDS catalog \citep{kubo_2018} but outside the PRIMER-UDS field are flagged as ``SXDS''. Those in the COSMOS field but outside the COSMOS-Web region are flagged as ``COSMOS''. Sources from the XMM-LSS catalog \citep{Desprez_2023} that are not included in the SXDS or PRIMER-UDS catalogs are flagged as ``XMM-LSS''.}
    \tablefoottext{b}{Source name in the references.}    
    \tablefoottext{c}{Spectroscopic confirmation was previously reported in these works: 1=\citet{tanaka_2019}, 2=\citet{valentino_2020a}, 3=\citet{schreiber_2018c}, 4=\citet{nanayakkara_2025}, 5=\citet{forrest_2020b} , 6=\citet{glazebrook_2017}, 7=\citet{schreiber_2018b}.}
    \tablefoottext{d}{The exposure time for the spectra taken in the DeepDive program.}
    \tablefoottext{e}{JWST NIRCam filter identifiers: Wide (W): 0.9 = F090W; 1.15 = F115W; 1.5 =F150W; 2 = F200W; 2.7 = F277W; 3.5 = F356W; 4.4 = F444W; Medium (M): 4.1 = F410M.}
    \tablefoottext{f}{HST filter identifiers: 0.44 = ACS/F435W; 0.6 = ACS/F606W; 0.8 = ACS/F814W; 1.25 = WFC3/F125W; 1.4 = WFC3/F140W; 1.6 = WFC3/F160W.}
    }
    \label{tab:photoinfo}
\end{table*}

\section{Comparison between the PSF-matched and the total photometry from DJA}
\begin{figure}
    \centering
    \includegraphics[width=0.8\linewidth]{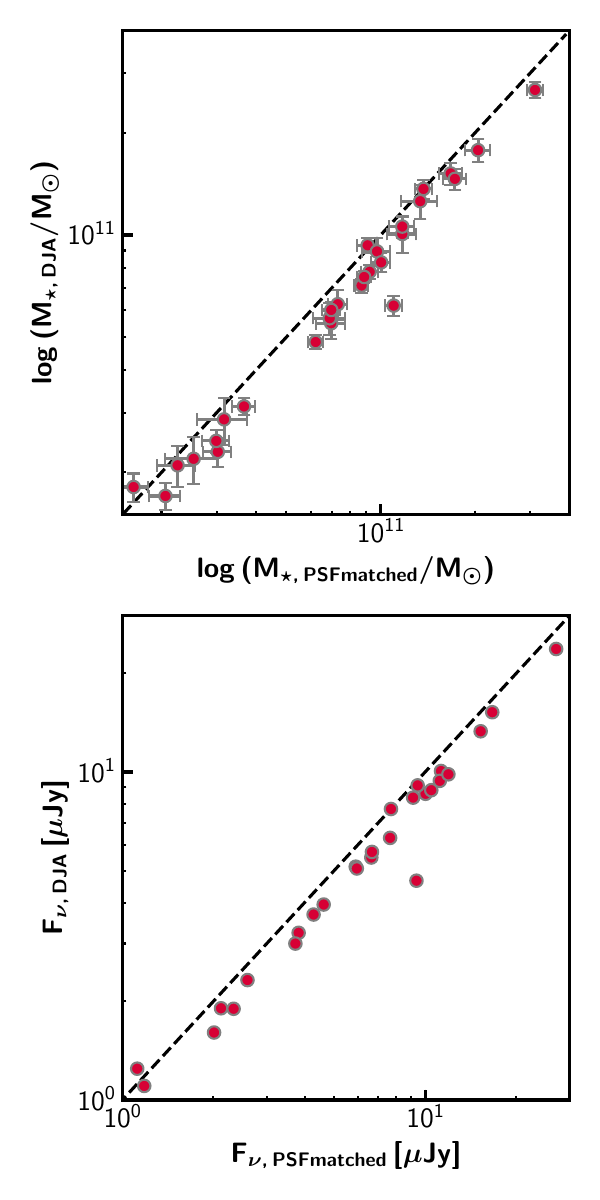}
    \caption{Top panel: Comparison between the stellar mass estimates based on the photometry extracted from the PSF-matched images and those available in DJA photometry. The quiescent galaxies observed in the DeepDive program are shown. Bottom panel: Comparison between the flux density in the F444W band from the two sets of photometric datasets in the top panel.}
    \label{fig:Mstar-comp}
\end{figure}

In Figure \ref{fig:Mstar-comp}, we compare the stellar mass and F444W flux density based on the photometry from the PSF-matched images with those from the photometry in DJA. 

\section{Flux correction functions for grating spectra in the extended wavelength range}
\label{app:Extendedcorr}
\begin{figure*}
    \centering
    \includegraphics[width=1\linewidth]{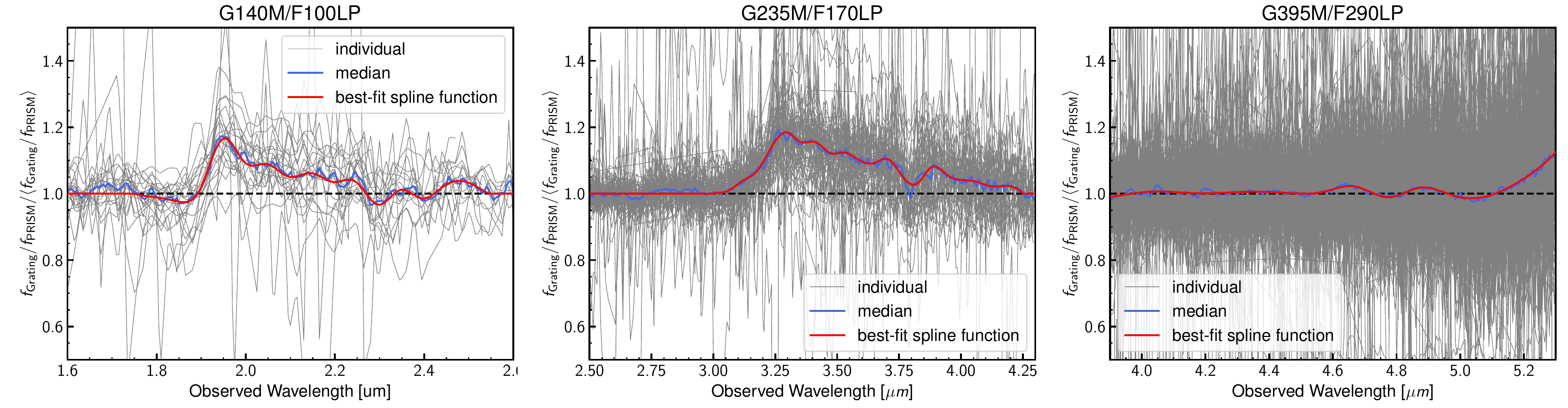}
    \caption{Ratio between the G140M/F100LP (left), G235M/F170LP (middle), and G395M/F290LP (right) spectra and PRISM/CLEAR spectra as a function of wavelength. The thin gray lines represent individual sources, while the blue line indicates their median. The best-fit spline function to the median ratio is shown in red and used as the correction function.}
    \label{fig:extendedspec_corr}
\end{figure*}

Figure~\ref{fig:extendedspec_corr} shows the ratios between the grating and prism spectra for G140M/F100LP, G235M/F170LP, and G395M/F290LP, along with their best-fit spline functions (derived from the median of each sample) used for the second-order, wavelength-dependent correction. The objects used to derive the correction are selected from the DeepDive and DJA catalogs, requiring coverage with both grating and prism spectroscopy. The sample includes all types of sources, not limited to quiescent galaxies. In total, 25, 78, and 393 spectra were analyzed for G140M/F100LP, G235M/F170LP, and G395M/F290LP, respectively. Significant offsets are seen between the grating and prism v4 spectra beyond the nominal wavelength ranges, specifically over $1.9\,{\rm \mu m}<\lambda_{\rm rest}<2.3\,{\rm \mu m}$ and $3.05\,{\rm \mu m}<\lambda_{\rm rest}<4.2\,{\rm \mu m}$ for G140M/F100LP and G235M/F170LP, respectively. These offsets likely arise from the lack of dedicated calibration spectra. For G395M/F290LP, the correction is largely flat, except near the spectral edge ($\lambda>5.2\,\mu$m).

\section{The configuration of the SED fitting with {\sc Bagpipes}}
Table \ref{tab:priors} summarizes the free parameters and priors of the SED fitting with {\sc Bagpipes} (Section~\ref{sec:SEDfitting}).
\begin{table}[]
    \centering
    \caption{Free parameters and priors for the photometric modeling with {\sc Bagpipes} .}
    \begin{tabular}{lcc}
    \toprule
    \toprule
        Free parameter & Prior & Limits \\
    \midrule
        $\mathrm{log}(M_{\rm formed}/M_\star)$ & Uniform & (0, 13)\\
        $A_{\rm V}/{\rm mag}$ &                            Uniform & (0,  4)\\
        $Z / Z_\odot$ &                          Logarithmic & (0.2,  5)\\
        $\tau / \mathrm{Gyr}$ &                  Uniform & (0.1,  $t(z_{\rm obs})$)\tablefootmark{a}\\
        $\alpha$, $\beta$ &                  Logarithmic & (10$^{-2}$, 10$^{3}$)\\
    \bottomrule
    \end{tabular}
    \tablefoot{
    \tablefoottext{a}{$t(z_{\rm obs})$ is the time since the Big Bang at the redshift of the sources.}
    }
    \label{tab:priors}
\end{table}

\section{List of JWST programs contributing to the archival compilation}\label{app:archivallist}
The archival grating spectra of quiescent galaxies in DJA have been collected as part of the following programs: PID\#1180, 1181 \citep[JADES, PI: D. Eisenstein,][]{eisenstein_2023}, PID\#1207 \citep[MIDIS, PI: G.Rieke,][]{ostlin_2025}, PID\#1215, 1286 \citep[JADES, PI: N. Luetzgendorf,][]{eisenstein_2023}, PID\#1671 \citep[PI: M.Maseda,][]{Maseda_2023}, PID\#1810 \citep[Blue Jay, PI: S.Belli,][]{davies_2024,park_2024}, PID\#1879 (PI: M.Curti), PID\#1914 \citep[AURORA, PI: A.Shapley,][]{Shapley_2025}, PID\#2110 \citep[SUSPENSE, PI: M. Kriek,][]{Slob_2024}, PID\#3215 (PI: D. Eisenstein), PID\#3543 \citep[EXCELS, PI: A. Carnall,][]{carnall_2024}, PID\#4233 \citep[RUBIES, PI: A. de Graaff,][]{degraaff_2024_rubies}, PID\#4446 (PI: B. Frye). Also, the PRISM/CLEAR spectra from the following programs are used for the flux correction: PID\#1180, 1181 (PI: D. Eisenstein), PID\#1212, 1215, 1286 (PI: N. Luetzgendorf), PID\#2565 \citep[PI: K. Glazebrook,][]{nanayakkara_2024,nanayakkara_2025}, PID\#3215  (PI: D. Eisenstein), PID\#4233 \citep[PI: A. de Graaff and G. Brammer,][]{degraaff_2024_rubies}, PID\#4446 (PI: B. Frye).

\section{Spectra and catalog of quiescent galaxies selected in the archival search}
\begin{figure*}
    \centering
    \includegraphics[width=0.8\linewidth]{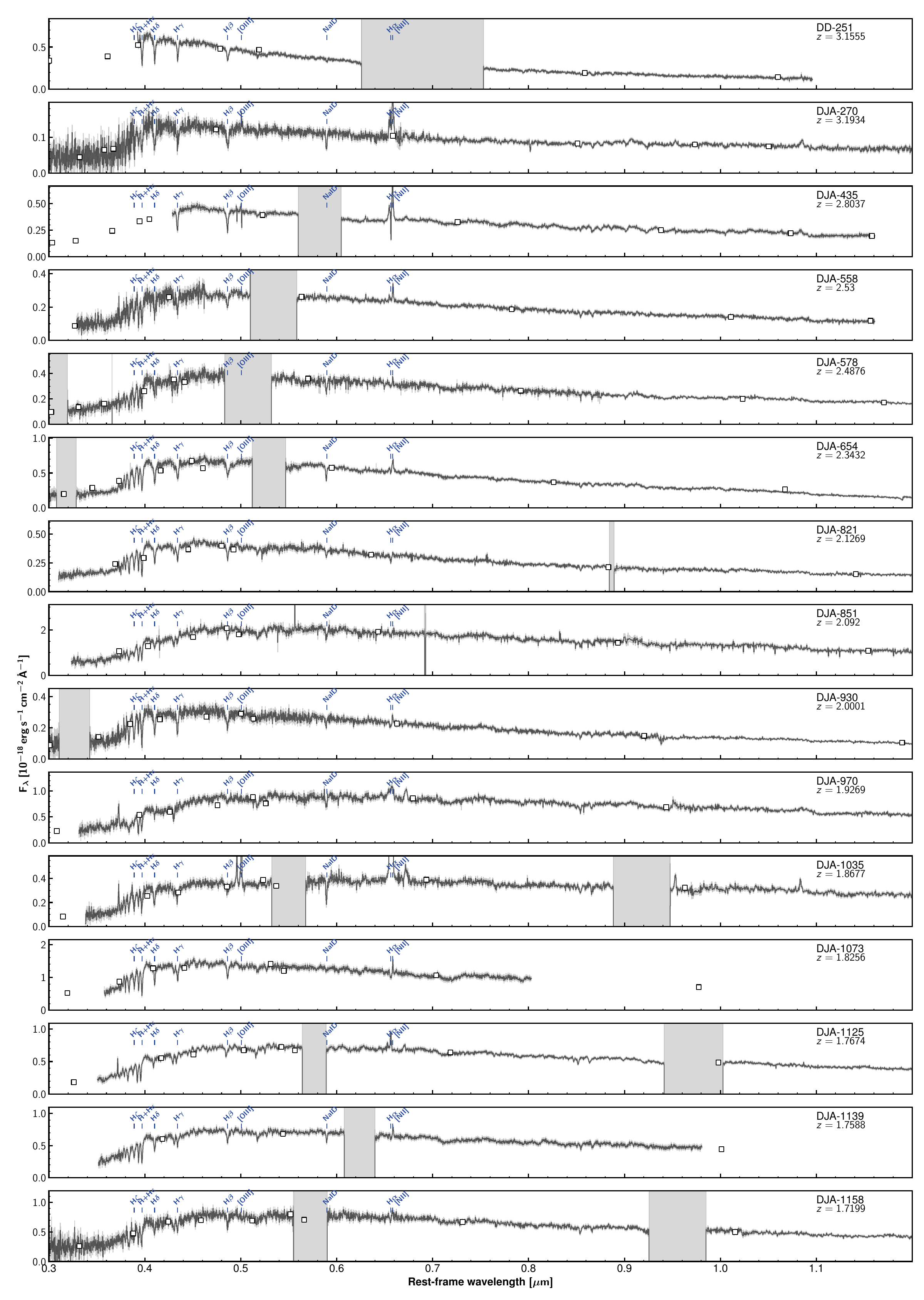}
    \caption{Examples of grating spectra of quiescent galaxies selected in Section \ref{sec:archive}. This subsample is composed of sources at $z>1.7$ whose spectra have a median $S/N>20$ on the stellar continuum. The solid line and squares correspond to their spectra and photometry, respectively.}
    \label{fig:archive-spectra}    
\end{figure*}

\begin{figure*}
    \centering
    \includegraphics[width=0.8\linewidth]{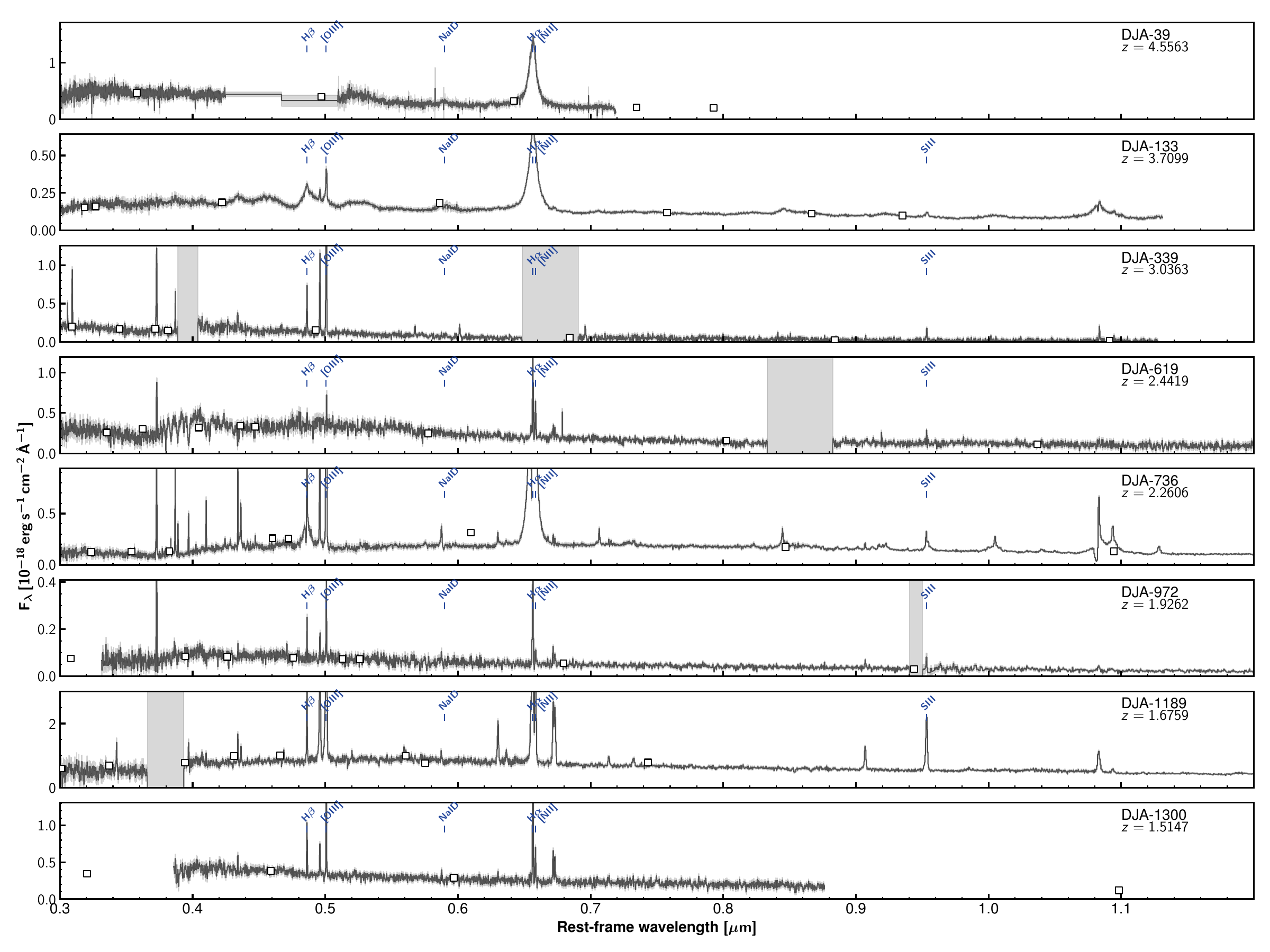}
    \caption{List of the spectra of the removed sources through visual inspection. The color and line codes are as in Figure \ref{fig:archive-spectra}.}
    \label{fig:archive-susspectra}    
\end{figure*}

Figure \ref{fig:archive-spectra} shows a subsample of high signal-to-noise ratio grating spectra of quiescent galaxies selected in Section~\ref{sec:archive}. The properties of the quiescent galaxies selected in Section~\ref{sec:archive} are listed in Table \ref{tab:archivalQG}. The full list is available online\footnote{\url{https://doi.org/10.5281/zenodo.18508154}}. For transparency and reference, the spectra of continuum-selected galaxies removed from the quiescent galaxies sample through the visual inspection are shown in Figure \ref{fig:archive-susspectra}. 

\begin{table*}[]
    \small
    \centering
    \caption{Properties of the quiescent galaxies selected from the archival search\tablefootmark{a}.}
    \begin{tabular}{lccccccccc}
    \toprule
    ID\tablefootmark{b} & (R.A., Decl.) & $z_{\rm spec}$ & $M_\star$\tablefootmark{c} & SFR\tablefootmark{c} & \dn\ \tablefootmark{d}& ${\rm QG_{D_n 4000}}$\tablefootmark{e} & ${\rm QG_{UVJ}}$\tablefootmark{f} &  ${\rm QG_{sSFR}}$\tablefootmark{g} & Ref.\tablefootmark{h}\\
     & & & [$10^{10}\, M_\odot$] & [$M_\odot\, {\rm yr^{-1}}$] & & & & &\\
    \midrule
DJA-27  & (214.915546, 52.949018) & 4.9000 & $8.20 \pm 0.52$ & $6.6 \pm 1.4$ & - & -1 & 1 & 1 & 1\\
DJA-36  & (34.365072, -5.148846)  & 4.6216 & $8.2  \pm 1.1$  & $0.000 \pm 0.035$ & $1.290 \pm 0.021$ & 1 & 1 & 1 & 2\\
DD-53   & (34.399676, -5.136348)  & 4.6193 & $10.0 \pm 1.2$ & $0.00 \pm 0.04$ & $1.398 \pm 0.021$ & 1 & 1 & 1 & 2\\
DJA-53  & (34.280543, -5.217151)  & 4.3599 & $5.82  \pm 0.41$  & $0.00 \pm 0.12$ & $1.568 \pm 0.033$ & 1 & 1 & 1& 2\\
DJA-65  & (189.265718, 62.168393) & 4.1438 & $2.45 \pm 0.17$ & $1.53 \pm 0.61$ & $1.249 \pm 0.046$ & 1 & 0 & 1 & 3, 4\\
DJA-71  & (171.823613, 42.469639) & 4.1065 & $4.77 \pm 0.54$ & $0.0 \pm 3.9$ & $1.239 \pm 0.016$ & 1 & 0 & 1 & 5, 6\\
DD-76   & (34.329342, -5.039926)  & 4.0125 & $2.9  \pm 0.5$  & $5.7 \pm 1.7$ & $1.110 \pm 0.047$ & 0 & 1 & 1& \\
DD-79   & (34.311943, -5.008798)  & 3.9933 & $1.8  \pm 0.2$  & $4.4 \pm 1.8$ & $1.373 \pm 0.070$ & 1 & 0 & 1 &\\
DD-82   & (34.316190, -5.051441)  & 3.9850 & $2.1  \pm 0.3$  & $16\pm 12$& $1.044 \pm 0.046$ & 0 & 1 & 0 &\\
DJA-91  & (34.360159, -5.153092)  & 3.9783 & $5.1  \pm 0.6$  & $0.00 \pm 0.87$ & $1.432 \pm 0.060$ & 1 & 1 & 1& 2\\
DJA-92  & (34.242590, -5.143077)  & 3.9746 & $3.3  \pm 0.3$  & $0.0 \pm 1.1$ & $1.104 \pm 0.015$ & 0 & 0 & 1 & 
 2\\
DD-126  & (150.117783, 2.386857)  & 3.7514 & $1.7  \pm 0.2$  & $2.0 \pm 1.2$ & $1.171 \pm 0.029$ & 0 & 0 & 1 &\\
DD-129  & (150.066682, 2.382370)  & 3.7309 & $5.7  \pm 0.6$  & $0.00 \pm 0.88$ & $1.173 \pm 0.053$ & 0 & 1 & 1& 7\\
DJA-134 & (34.290443, -5.262081)  & 3.7013 & $2.7  \pm 0.2$  & $0.00 \pm 0.19$ & - & -1 & 1 & 1 & 2, 7\\
DD-144  & (149.945536, 2.111363)  & 3.6511 & $2.2  \pm 0.3$  & $1 \pm 14$ & $1.138 \pm 0.042$ & 0 & 1 & 1 &\\
DJA-156 & (53.196910, -27.760528) & 3.6177 & $4.2 \pm 0.2$ & $0.000 \pm 0.049$ & $1.179 \pm 0.032$ & 0 & 1 & 1 &3\\
DJA-163 & (53.120018, -27.852025) & 3.5851 & $1.9 \pm 0.2$ & $5.1 \pm 1.4$ & $1.184 \pm 0.042$ & 0 & 0 & 1 &\\
DD-165  & (149.955262, 2.088817)  & 3.5100 & $2.3  \pm 0.3$  & $0.01 \pm 0.72$ & $1.252 \pm 0.072$ & 1 & 1 & 1 &\\
DJA-176 & (53.082581, -27.866803) & 3.4858 & $2.7 \pm 0.1$& $0.000 \pm 0.014$  & $1.169 \pm 0.030$ & 0 & 1 & 1 &\\
DD-185  & (214.866046, 52.884089) & 3.4433 & $6.2 \pm 0.4$ & $0.0 \pm 0.8$ & $1.331 \pm 0.064$ & 1 & 1 & 1 & 8\\
DD-229  & (214.895613, 52.856498) & 3.2518 & $7.5 \pm 0.3$ & $0.00 \pm 0.02$ & $1.254 \pm 0.015$ & 1 & 0 & 1 & 7, 9\\
DJA-245 & (34.344085, -5.239474)  & 3.2381 & $2.8  \pm 0.3$  & $1.37 \pm 1.54$ & $1.421 \pm 0.043$ & 1 & 1 & 1 &2\\
DD-236  & (34.427658, -5.152416)  & 3.2268 & $12 \pm 1$ & $0.000 \pm 0.027$ & $1.235 \pm 0.113$ & 1 & 0 & 1& \\
DJA-269 & (34.255875, -5.233825)  & 3.1949 & $14 \pm 1$ & $0.000 \pm 0.060$ & $1.747 \pm 0.065$ & 1 & 1 & 1 & 2, 10\\
DJA-270 & (34.258890, -5.232293)  & 3.1934 & $5.5  \pm 0.5$  & $0.0 \pm 3.9$ & $1.490 \pm 0.047$ & 1 & 1 & 1 & 2, 7\\
DD-251  & (34.774841, -5.354522)  & 3.1555 & $6.3  \pm 0.6$  & $10\pm 25$& -                 & -1 & 0 & 1 &\\
DD-256  & (34.290242, -5.038906)  & 3.1326 & $2.5  \pm 0.2$  & $0.55 \pm 0.92$ & -                 & -1 & 0 & 1 &\\
DD-257  & (34.291081, -5.038119)  & 3.1307 & $18 \pm 1$ & $1.4 \pm 2.1$ & -                 & -1 & 1 & 1 &\\
DJA-314 & (150.097314, 2.260115)  & 3.0937 & $0.54  \pm 0.03$  & $0.00 \pm 0.27$ & $1.111 \pm 0.049$ & 0 & 0 & 1 &\\
DJA-321 & (34.364641, -5.250241)  & 3.0800 & $3.5  \pm 0.5$  & $0.0000 \pm 0.0046$ & $1.480 \pm 0.055$ & 1 & 1 & 1& 2\\
DJA-329 & (53.165314, -27.814140) & 3.0638 & $12.8 \pm 0.9$& $2.7 \pm 2.9$ & $1.161 \pm 0.017$ & 0 & 1 & 1 & 3, 11\\
DD-270  & (34.785875, -5.357320)  & 3.0153 & $6.0  \pm 0.3$  & $0.5 \pm 0.5$ & -                 & -1 & 1 & 1& \\
\bottomrule
\end{tabular}
    \tablefoot{    
    \tablefoottext{a}{The complete catalog is available in machine-readable format.}
    \tablefoottext{b}{IDs starting with {\tt DJA} represent objects from the Dawn JWST Archive; those starting with {\tt DD} represent secondary quiescent targets from the DeepDive program.}
    \tablefoottext{c}{If the photometry is not available, the stellar mass and SFR estimates are not listed.}
    \tablefoottext{d}{If the spectra do not cover the $0.38\, {\rm \mu m}<\lambda_{\rm rest}<0.41\, {\rm \mu m}$ range, the \dn\ estimate is not listed.}
    \tablefoottext{e, f, g}{These flags are set to $1$ if a source satisfies the \dn, $UVJ$, and sSFR selection criteria for quiescent galaxies, respectively. If sources do not have \dn\ measurements or photometry, their flags are set to $-1$.}
    \tablefoottext{h}{List of references reporting the original spectroscopic confirmation: 1=\citet{deGraaff_2024}, 2=\citet{carnall_2024}, 3=\citet{baker_2024}, 4=\citet{kokorev_2024}, 5=\citet{wu_2024}, 6=\citet{valentino_2025} , 7=\citet{nanayakkara_2025}, 8=\citet{ito_2025b}, 9=\citet{schreiber_2018c} , 10=\citet{glazebrook_2024}, 11=\citet{d'eugenio_2023}.}
    }
    \label{tab:archivalQG}
\end{table*}

\end{appendix}
\end{document}